\documentclass[dvipsnames]{elsarticle}
\usepackage{siunitx}
\usepackage{hyperref}
\usepackage{graphicx}  
\usepackage{dcolumn}   
\usepackage{bm}        
\usepackage{amssymb}   
\usepackage{amsfonts}
\usepackage{graphics}
\usepackage{grffile}   
\usepackage{epsfig}
\usepackage[usenames]{color}
\usepackage[normalem]{ulem} 
\usepackage{color}
\usepackage[utf8]{inputenc}
\usepackage[T1]{fontenc}
\usepackage{amsmath}
\usepackage{enumitem}

\usepackage{subcaption}
\usepackage{adjustbox}

\usepackage{lineno}

\DeclareSIUnit\ele{e^{\text{-}}}
\DeclareSIUnit{\NIEL}{\text{$1\ \text{MeV}\ \text{n}_{\textup{eq}}/\text{cm}^2$}}
\DeclareUnicodeCharacter{2212}{-}

\usepackage{lineno}

\usepackage{xcolor}
\usepackage{tikz}
\usepackage{environ}
\makeatletter
    \newsavebox{\measure@tikzpicture}
        \NewEnviron{scaletikzpicturetowidth}[1]{%
            \def\tikz@width{#1}%
            \def\tikzscale{1}\begin{lrbox}{\measure@tikzpicture}%
            \BODY
            \end{lrbox}%
            \pgfmathparse{#1/\wd\measure@tikzpicture}%
            \edef\tikzscale{\pgfmathresult}%
            \BODY
        }
\makeatother
\usetikzlibrary{math}
\usetikzlibrary{calc}
\usetikzlibrary{decorations.pathreplacing, decorations.markings}
\usetikzlibrary{patterns, arrows, arrows.meta}

\usepackage{imakeidx}
\makeindex[columns=3, title=Alphabetical Index, intoc]
\makeindex

\journal{Nucl. Instrum. Methods Phys. Res. A}

\begin{document}
\begin{frontmatter}
\title{Influence of Radiation and AC Coupling on Time Performance of Analog Pixels Test Structures in 65 nm CMOS technology}


\author[CERN]{Gianluca Aglieri Rinella}
\author[unito,infnTO]{Luca Aglietta}
\author[infnTS]{Matias Antonelli}
\author[uniba,infnBA]{Francesco Barile}
\author[infnTO]{Franco Benotto}
\author[unito,infnTO]{Stefania Maria Beolè}
\author[unito,infnTO]{Elena Botta}
\author[poliBA,infnBA]{Giuseppe Eugenio Bruno}
\author[infnBA]{Giovanni Francesco Ciani}
\author[uniba,infnBA]{Domenico Colella}
\author[uniba,infnBA]{Angelo Colelli}
\author[units,infnTS]{Giacomo Contin}
\author[infnBA]{Giuseppe De Robertis}
\author[infnTO]{Floarea Dumitrache}
\author[infnBA]{Domenico Elia}
\author[infnTO]{Chiara Ferrero}
\author[nikhef]{Martin Fransen}
\author[nikhef,utu]{Alessandro Grelli}
\author[CERN]{Hartmut Hillemanns}
\author[nikhef,uva]{Isis Hobus}
\author[CERN]{Alex Kluge}
\author[infnBA]{Shyam Kumar}
\author[CERN,IPHC]{Corentin Lemoine}
\author[infnBA]{Francesco Licciulli}
\author[unito,infnTO]{Bong-Hwi Lim}
\author[infnBA]{Flavio Loddo}
\author[nikhef,uva]{Esther Mwetaminwa M'Bilo}
\author[CERN]{Magnus Mager}
\author[unica,infnCA]{Davide Marras}
\author[CERN]{Paolo Martinengo}
\author[infnBA]{Cosimo Pastore}
\author[infnBA,uniJAM]{Rajendra Nath Patra}
\author[unito,infnTO]{Stefania Perciballi}
\author[CERN]{Francesco Piro}
\author[infnTO]{Francesco Prino}
\author[infnTO,UPO]{Luciano Ramello}
\author[CERN]{Felix Reidt}
\author[nikhef,uva]{Roberto Russo}
\author[CERN]{Valerio Sarritzu}
\author[unito,infnTO]{Umberto Savino\corref{cor1}}
\author[IPHC]{Serhiy Senyukov}
\author[infnTO,UPO]{Mario Sitta}
\author[CERN]{Walter Snoeys}
\author[nikhef,uva]{Jory Sonneveld}
\author[CERN]{Miljenko Suljic}
\author[poliBA,infnBA]{Triloki Triloki}
\author[infnCA,unica]{Gianluca Usai}
\author[nikhef,uva]{Håkan Wennlöf}

\affiliation[CERN]{organization={European Organisation for Nuclear Research (CERN)},
city={Geneva},
country={Switzerland}
}
\affiliation[unito]{organization={University of Torino - Department of Physics},
city={Torino},
country={Italy}
}
\affiliation[infnTO]{organization={INFN, Sezione di Torino},
city={Torino},
country={Italy}
}
\affiliation[infnTS]{organization={INFN, Sezione di Trieste},
city={Trieste},
country={Italy}
}
\affiliation[uniba]{organization={University of Bari - Department of Physics},
city={Bari},
country={Italy}
}
\affiliation[infnBA]{organization={INFN, Sezione di Bari},
city={Bari},
country={Italy}
}
\affiliation[poliBA]{organization={Polytechnic of Bari - Department of Physics DIF}, 
city={Bari},
country={Italy}
}
\affiliation[units]{organization={University of Trieste},
city={Trieste},
country={Italy}
}
\affiliation[nikhef]{organization={National Institute for Subatomic Physics (Nikhef)},
city={Amsterdam},
country={Netherlands}
}
\affiliation[uva]{organization={University of Amsterdam},
city={Amsterdam},
country={Netherlands}
}
\affiliation[utu]{organization={Utrecht University}, city={Utrecht}, country={Netherlands}}
\affiliation[IPHC]{organization={Université de Strasbourg, CNRS, IPHC UMR 7178}, 
city={Strasbourg}, 
country={France}
}
\affiliation[unica]{organization={University of Cagliari},
city={Cagliari},
country={Italy}
}
\affiliation[infnCA]{organization={INFN, Sezione di Cagliari},
city={Cagliari},
country={Italy}
}
\affiliation[uniJAM]{organization={Department of Physics, University of Jammu}, 
city={Jammu}, 
country={India}
}
\affiliation[UPO]{organization={Università del Piemonte
Orientale},
city={Vercelli},
country={Italy}
}

\cortext[cor1]{Corresponding author: umberto.savino@unito.it}

\begin{abstract}
Monolithic Active Pixel Sensors (MAPS) in advanced CMOS imaging technologies are key to next-generation tracking systems for high-energy physics, where radiation hardness and precise vertex reconstruction are essential. 
As part of the ALICE ITS3 R\&D program in synergy with the CERN R\&D, we evaluated the performance of the Analog Pixel Test Structures (APTS) fabricated in the TPSCo 65 nm CMOS imaging process. The prototypes employ \SI{10}{\um} pitch pixels with a fast operational amplifier–based buffering stage at the output, enabling direct characterization of intrinsic sensor response. 
Beam tests with minimum ionizing particles assessed the timing and charge collection of DC- and AC-coupled designs, including devices exposed to \SI{d14}{\NIEL} and \SI{d15}{\NIEL} non ioninsing energy loss. DC-coupled sensors demonstrated stable performance, maintaining time resolution lower than \SI{70}{ps} and $>99$\% detection efficiency up to \SI{d15}{\NIEL}.

AC-coupled sensors demonstrated a wide operational margin, with efficiencies above 99\% for clusterization thresholds below \SI{150}{e^-}. Even though the AC coupling allows higher reverse bias than DC-coupled sensors, the reduced signal amplitude lowers the signal-to-noise ratio, increasing the jitter contribution. At high reverse bias, the AC-coupled sensors achieve time resolutions comparable to the DC-coupled version, demonstrating the viability of both approaches. These results also suggest that combining the low capacitance of DC-coupled designs with the high-bias capability of AC coupling could further enhance time resolution.

These results confirm the suitability of 65 nm MAPS for future collider detectors requiring high radiation tolerance, efficiency, and timing precision.

\end{abstract}

\begin{keyword}
Monolithic Active Pixel Sensors \sep Solid state detectors \sep Silicon \sep CMOS \sep Particle detection \sep Test beam
\end{keyword}

\end{frontmatter}

%
%
\tikzset{
	beamarrow/.style={
		decoration={
			markings,mark=at position 1 with 
			{\arrow[scale=2,>=stealth]{>}}
		},postaction={decorate}
	}
}

\tikzset{
	pics/.cd,
	vector out/.style={
		code={
		\draw[#1, thick] (0,0)  circle (0.15) (45:0.15) -- (225:0.15) (135:0.15) -- (315:0.15);
		}
	}
}
\tikzset{
	pics/.cd,
	vector in/.style={
		code={
		\draw[#1, thick] (0,0)  circle (0.15);
		 \fill[#1] (0,0)  circle (.05);
		 }
	}
}
\tikzset{
	global scale/.style={
		scale=#1,
		every node/.style={scale=#1}
	}
}
\def\centerarc[#1] (#2)(#3:#4:#5) 
	 { \draw[#1] ($(#2)+({#5*cos(#3)},{#5*sin(#3)})$) arc (#3:#4:#5); }
\section{Introduction}
The development of high-granularity, low-material tracking systems is a key requirement of modern particle physics experiments. 
As collider luminosities and data rates continue to increase~\cite{Aberle:2749422}, innovative sensor technologies are required to maintain excellent spatial and temporal resolution, while minimizing material budget. 
In this context, Monolithic Active Pixel Sensors (MAPS) have emerged as a compelling solution due to their fine segmentation, low power consumption, and integration of sensor and readout circuitry on the same silicon die~\cite{CONTIN201860, REIDT2022166632}.

A pioneering implementation of large-scale MAPS in a high-energy physics environment is the Inner Tracking System (ITS2) of the ALICE (A Large Ion Collider Experiment) detector at the Large Hadron Collider (LHC). Commissioned in 2021, covering a surface of approximately \SI{10}{\square\meter}, ITS2 is the largest application of MAPS in a collider experiment~\cite{REIDT2022166632, Abelev_2014, Acharya_2024}. 
This marks a significant technological milestone, establishing MAPS as a viable option for next-generation tracking systems.

To further enhance ALICE vertexing performance, the upgrade of the ITS~\cite{loi,doi:10.7566/JPSCP.34.010011, The:2890181} in LS3, namley the ITS3, aims to develop a new generation of MAPS using the Tower Partners Semiconductor Co. (TPSCo) \SI{65}{\nm} CMOS imaging process, TPSCo 65 nm ISC~\cite{towerjazz}. This technology offers enhanced radiation tolerance, reduced pixel pitch, and faster signal processing capabilities, compared to previously used technologies. As part of the ITS3 and EP R\&D programs, an Analog Pixel Test Structure (APTS) was designed in several variants as a test chip aimed at general sensor characterization. Due to its fast output stage, especially for the variant with the operational amplifier (OA)-based output stage, it can also be used to characterize the sensor timing performance.

The APTS prototype includes a $4\times 4$ matrix of \SI{10}{\um} square pixels. It has been implemented with several sensor variants. In the one studied here, each pixel features a fast operational amplifier (OA)-based buffering stage, and it employs a n-type implantation in the epitaxial layer to optimize charge collection~\cite{Munker_2019}. 
By providing direct access to the analog output of individual pixels, with sufficient bandwidth, the APTS allows detailed characterization of the charge collection properties of the sensor. 
Previous beam test studies have demonstrated excellent performance, combining a time resolution of \SI{63}{\ps}, spatial resolution of \SI{2}{\um}, and detection efficiency exceeding 99\% for minimum ionizing particles \cite{opamp_2025}.

In this work, we present results from a test beam campaign aimed at evaluating the radiation hardness of the APTS-OA DC-coupled sensors up to a fluence of \SI{d15}{\NIEL}, and compare the performance of non-irradiated DC\cite{opamp_2025} and AC-coupled sensors. 
The APTS-OA time resolution and charge collection efficiency are assessed for all the above sensors.
\section{The APTS-OA}
The APTS-OA consists of a $4 \times 4$ matrix of \SI{10}{\um}-pitch pixels, each directly buffered to output pads for analog readout and operated via external reference currents and voltages. 
Dummy pixels surround the matrix to mitigate edge-related electric field distortions. 
The radiation hardness of the DC-coupled design was studied through a non-ionizing energy loss (NIEL) irradiation campaign with neutrons at JSI Ljubljana.
APTS-OA sensors were irradiated at fluences of \SI{d14}{\NIEL} and \SI{d15}{\NIEL}.

\subsection{Sensor variants}

\begin{figure}[ht]
    \centering
    \includegraphics[width=0.6\linewidth]{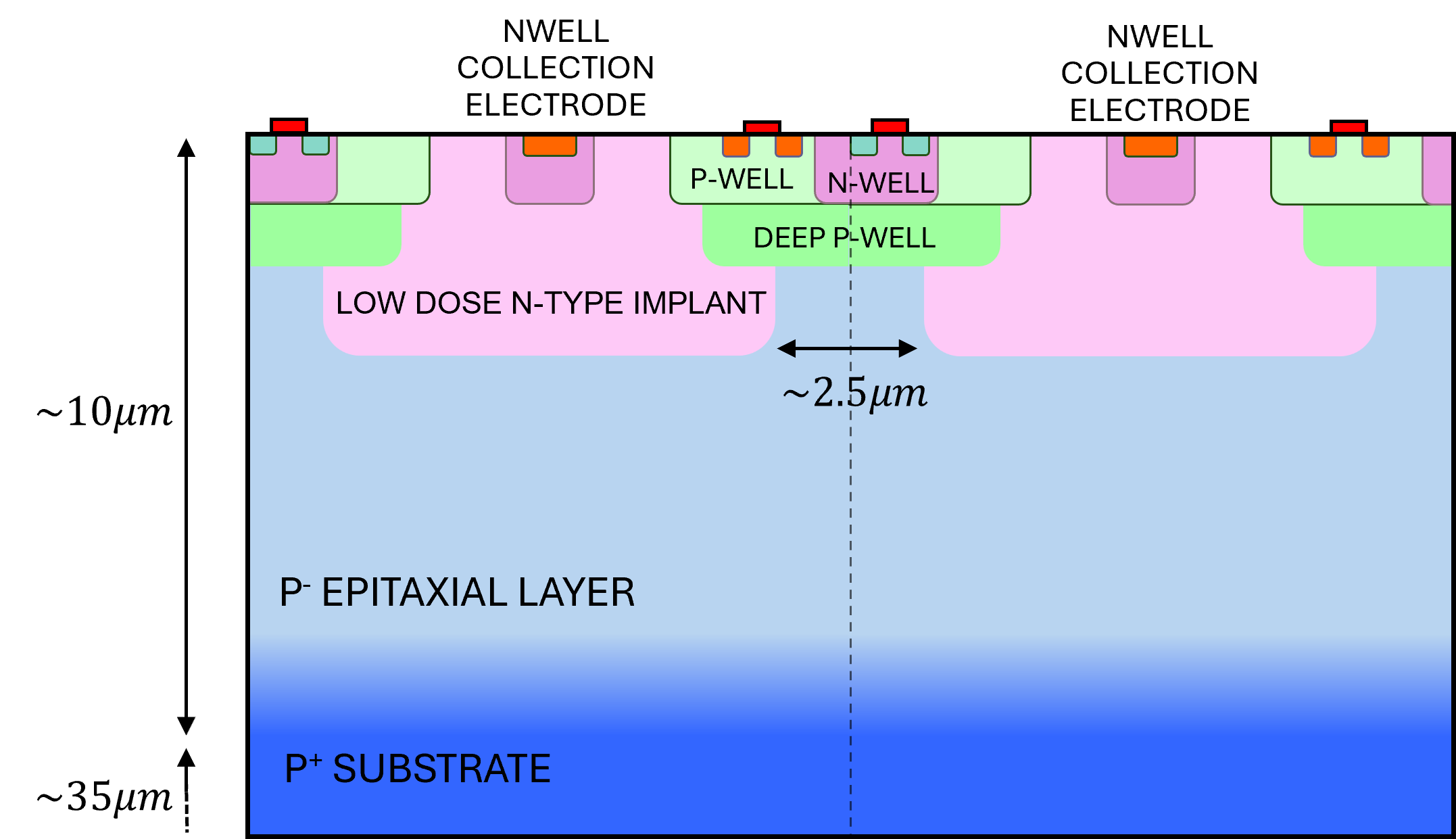}
    \caption{Schematic cross section of two pixels in the \textit{modified with gap} process.}
    \label{fig:cross_section}
\end{figure}
The APTS was fabricated using three sensor process variants, derived from a standard imaging CMOS design and progressively modified to enhance charge collection speed and reduce charge sharing~\cite{AGLIERIRINELLA2023168589,APTSSF_2023,SNOEYS201790}. 
In this study, we focus on the \textit{modified with gap} variant, which features a small n-well collection electrode (about \SI{1}{\um} diameter) extended by a low dose deep n-type implant in a high-resistivity p-type epitaxial layer. A gap in the deep n-type implant near the pixel edges creates a lateral electric field, accelerating charge collection, improving timing performance and radiation hardness. A deep p-well allows full CMOS circuitry in the pixel matrix without compromising charge collection efficiency.

Two versions of coupling between the sensor and the front end are available. In the DC-coupled version, already presented in detail in \cite{opamp_2025}, the sensor is directly connected to the input of the front-end. In that case the potential of the collection electrode is dictated by the operating point of the front end, with an upper limit of \SI{1.2}{\volt}. Further reverse biasing of the sensor has to be achieved by lowering the substrate voltage to enhance charge collection. Since the p-wells containing transistors inside the matrix are exposed to the same potential as the substrate, this version is limited to a maximum of \SI{6}{\volt} of potential difference to avoid breakdown of the shallow source and drain junctions. Consequently, the substrate bias is limited to about \SI{4.8}{\volt}. However, the small collection electrode leads to a low capacitance of the input node, which results in a significant signal despite this limited reverse bias. 

\begin{figure}[ht]
    \centering
    \includegraphics[width=\linewidth]{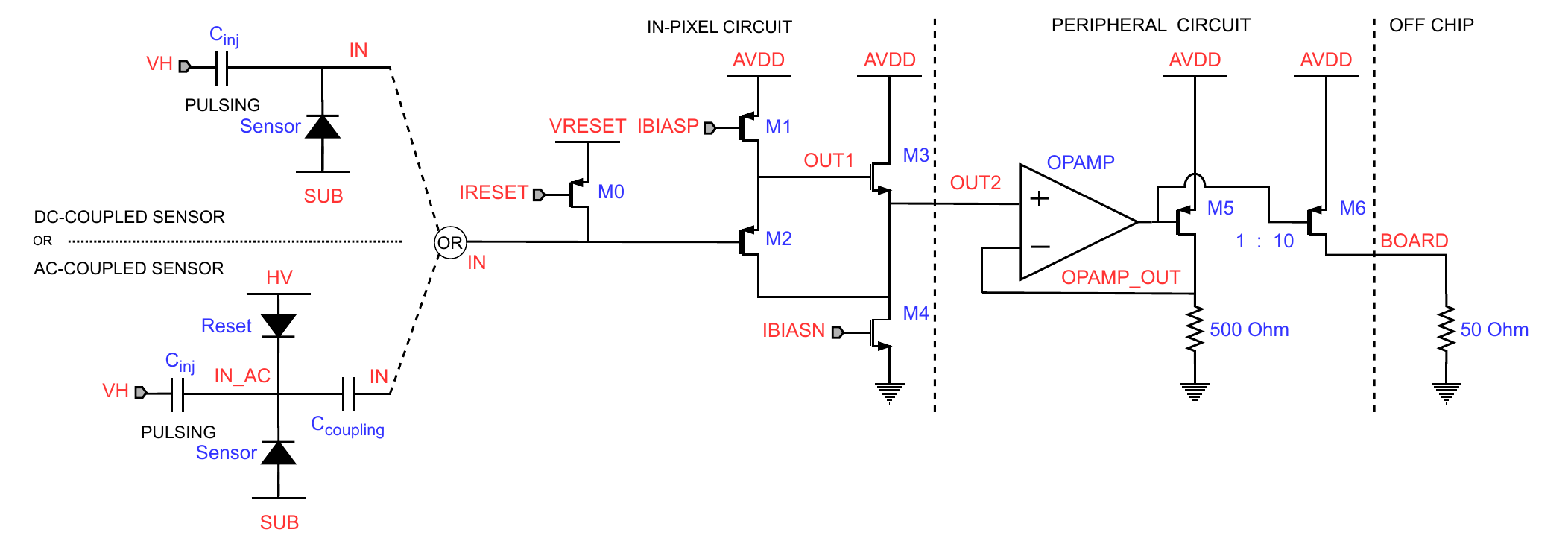}
    \caption{Schematic of the front-end of the DC-coupled and AC-coupled APTS-OA.}
    \label{fig:front-end}
\end{figure}
To explore higher reverse sensor biases, an AC-coupled design was also manufactured. The schematics of both DC and AC-coupled versions are visible in Figure~\ref{fig:front-end}. The AC-coupled version separates the collection electrode node ($IN\_AC$) and the front-end input ($IN$) with a metal-to-metal capacitance $C_\text{coupling} \approx $ \SI{7.9}{\femto\farad} (simulated value). The collection electrode can then be biased at higher voltages (up to \SI{50}{\volt} from design),
but as no increase in signal amplitude was observed beyond \SI{18}{V} when testing up to \SI{37}{V}, all results in this paper are reported up to \SI{18}{V}. 
Since there is no DC path between the reset transistor $M0$ and the sensor, leakage current compensation and sensor reset are provided by the diode inside the collection electrode. Due to the coupling capacitance, the total capacitance at the input node is significantly higher than in the DC-coupled case and the same collected charge will lead to a smaller amplitude signal.
The front-end is identical for the two coupling versions and described in Section~\ref{sec:front-end}.

For consistency with previously published results on non-irradiated DC-coupled sensors \cite{opamp_2025}, the devices are operated at the same analog front-end settings.
Since the leakage current, estimated from the study of the signal reset \cite{APTSSF_2023}, is smaller than the default reset current (100 pA), the operating point was not changed for irradiated sensors.

\subsection{Analog circuit}
\label{sec:front-end}
The APTS-OA analog front-end provides a low-capacitance load for the sensor and a signal path designed with buffering for high-speed analog readout. 
Each pixel is connected to a source-follower buffer that drives a fast operational amplifier located at the matrix periphery. 
The front-end includes an in-pixel pulsing circuit for charge injection to support calibration. To maintain signal integrity, the output is matched to a \SI{50}{\ohm} load via a high-speed amplifier with a \SI{1.9}{\giga\hertz} unity-gain bandwidth. All bias voltages and currents are supplied externally, enabling fine control of the operating point
, which was experimentally optimized for minimal timing jitter. The estimated power consumption is \SI{150}{\uW} in-pixel and \SI{3.1}{\mW} in the periphery per pixel~\cite{opamp_2025}.

\subsection{Readout system}
\begin{figure}
    \centering
    \includegraphics[width=0.9\linewidth]{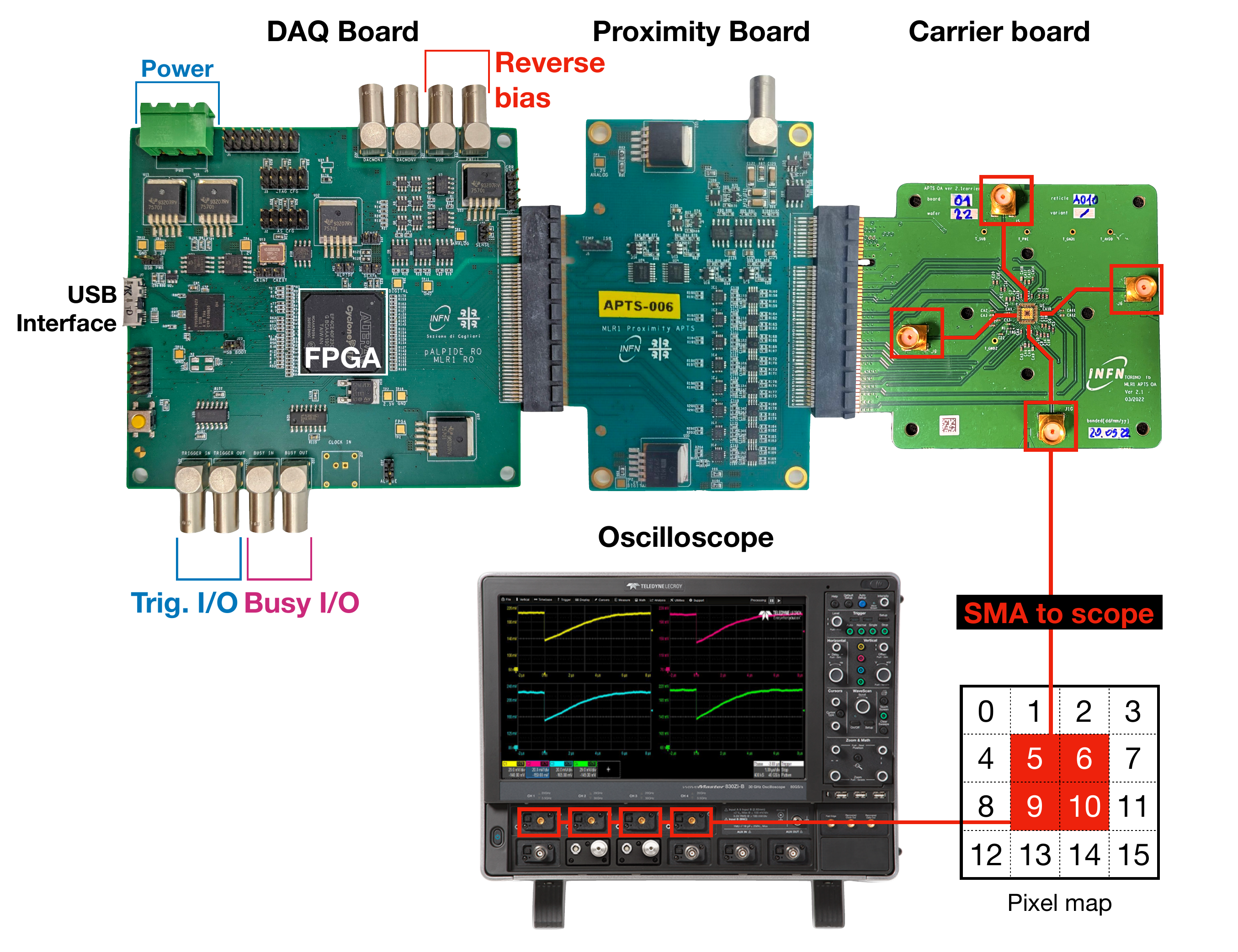}
    \caption{APTS-OA power, control, and readout setup from \cite{opamp_2025}.}
    \label{fig:OA-setup}
\end{figure}
The APTS-OA test setup (see Figure~\ref{fig:OA-setup}) consists of a \SI{5}{\volt} FPGA-based DAQ board (USB-controlled via PC), a proximity board providing power, biases, and ADC readout, a carrier board to which the sensor is mounted and wire-bonded, and an oscilloscope with high sampling rate (\SI{40}{GS/s}) and high bandwidth (\SI{13}{GHz}). Further details are available in Ref.~\cite{opamp_2025}.
The carrier board includes four high-bandwidth SMA connectors linked to the innermost pixels for oscilloscope readout; in this study, three of the four were used, as one channel was devoted to the time reference (cf. Section~\ref{sec:TBsetup}). 
The remaining 13 pixels (12 outer and one inner) were read out via \SI{4}{\mega\hertz} ADCs on the proximity board.

\section{Test Beam Measurements} 
The APTS-OA time resolution, detection efficiency and spatial resolution were characterized with a beam test conducted at the CERN-SPS H6 facility using positive \SI{120}{\giga\electronvolt/c} hadrons~\cite{Banerjee:2774716}.
 
\subsection{Test beam telescope}
\label{sec:TBsetup}
\begin{figure}[!ht]
    \centering
    \input{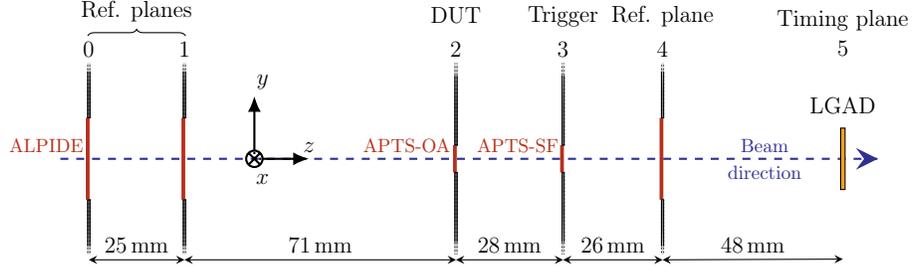}
    \caption{Schematic of the test beam setup showing each detector plane.}
    \label{fig:telescope}
\end{figure}
The setup, as illustrated in Figure \ref{fig:telescope}, consisted of six planes:
\begin{itemize}
    \item Planes 0, 1, and 4 are reference planes for track reconstruction. ALPIDE sensors \cite{ALPIDE-proceedings-3} were used at a threshold of \SI{100}{\ele} and a substrate bias of \SI{-3}{\volt}.
    \item Plane 2 is an APTS-OA as the device under test (DUT). DC coupled irradiated sensors and an AC-coupled sensor were characterized. The DUT was placed on xy moving stages for alignment before testing, and the temperature was stabilized at \SI{20}{^{\circ}C} using a chiller.
    \item Plane 3 is an APTS equipped with source-follower (SF) output buffer \cite{APTSSF_2023} used as trigger plane. The sensor has a pixel pitch of \SI{15}{\um} and is implemented with the modified with gap design. Like the DUT, this plane was mounted on xy moving stages. It was operated at a substrate bias of \SI{-1.2}{\volt} and all other biases were set to default 
    based on \cite{APTSSF_2023}.
    \item Plane 5 is a low gain avalanche detector (LGAD) produced by Fondazione Bruno Kessler \cite{Carnesecchi_2023} used as a time reference. It was operated at a bias voltage of \SI{110}{\volt}. 
\end{itemize}

A large number of events was collected to study the in-pixel contribution of the lateral electric field on charge collection. 
The full statistics collected for each sensor at each bias voltage is reported in \ref{apx:statistics}.

\subsection{Data reconstruction and analysis}
\begin{figure}[!ht]
    \centering
    \includegraphics[width=1.0\linewidth]{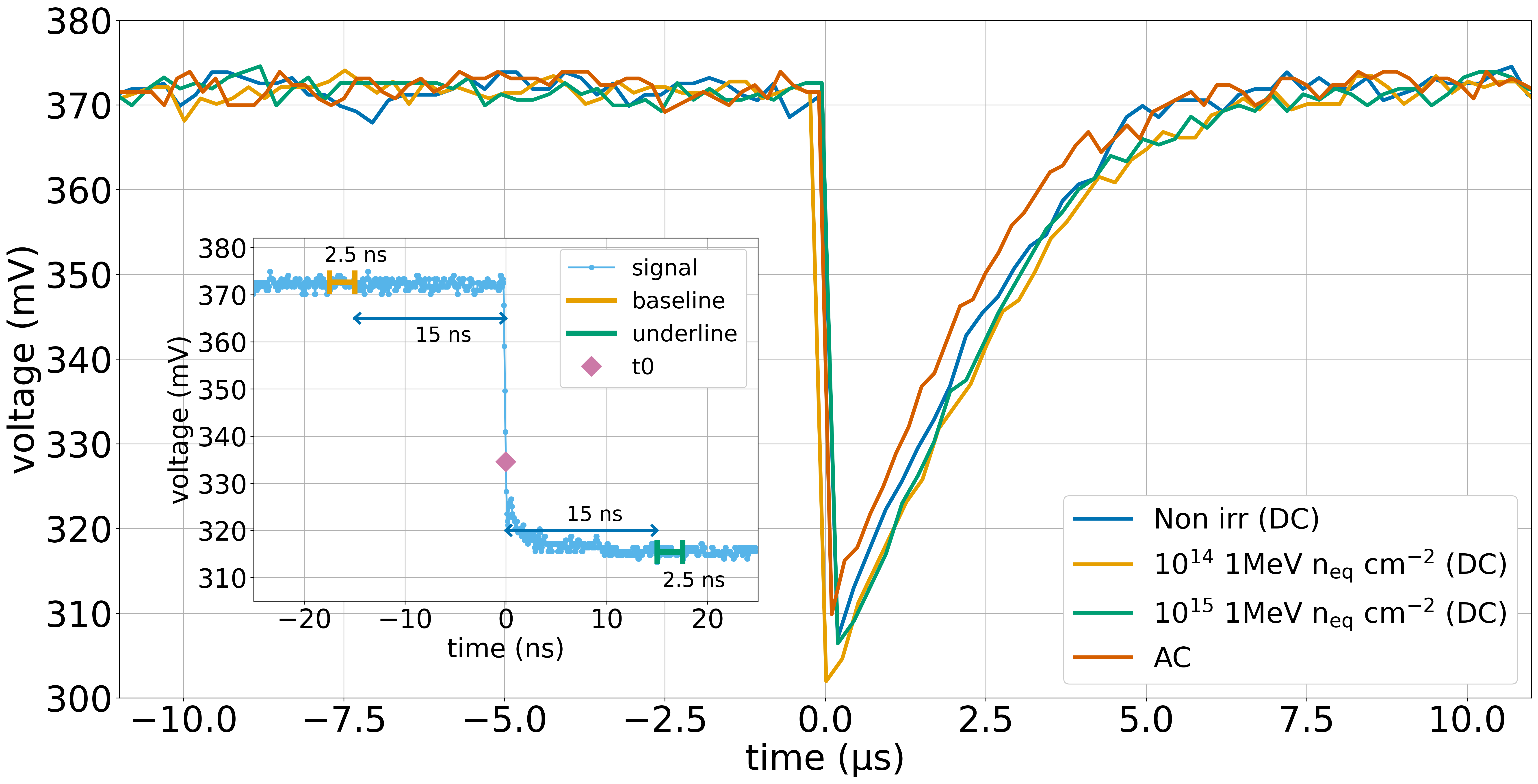}
    \caption{Typical signals measured after charge injection. The inset shows the data ranges used to define the baseline and the underline.}
    \label{fig:waveform}
\end{figure}

\paragraph{DUT signal extraction} 
The analog waveforms, shown in Figure~\ref{fig:waveform}, were analyzed offline following the method described in \cite{opamp_2025}. 
In short, the signal amplitude of the pixels read out via oscilloscope was extracted as the difference between the baseline and the underline, computed respectively as the average over a \SI{2.5}{\ns} window, taken \SI{15}{\ns} before and after the minimum of the waveform’s derivative, t0 in Figure~\ref{fig:waveform}.
For the pixels read out via ADC's, amplitude was defined as the difference between the minimum of the signal and the baseline taken four samples earlier, as described in Ref.~\cite{APTSSF_2023}.
\paragraph{DUT signal calibration}
To enable comparison across pixels and operating conditions, the measured signal amplitude was converted into collected charge (in electrons) using an X-ray calibration, following the method detailed in \cite{opamp_2025}. 
The spectrum of $^{55}$Fe, reported in \ref{apx:xray-spectra}, was fitted with a bimodal Gaussian to deconvolve overlapping $K_{\alpha}$ and $K_{\beta}$ peaks. 
The fitted mean of the $K_\alpha$ peak ($\mu_{peak}$) was then used for the conversion of millivolts to electrons via:
\begin{equation}
    Q = \frac{E_{\gamma}}{\mu_{peak}\cdot\epsilon}
\end{equation}
where $E_{\gamma}$ is the photon energy, assumed to be \SI{5.9}{\kilo \electronvolt}, and $\epsilon$ is the average energy required to generate an electron-hole pair in silicon, assumed to be \SI{3.6}{\electronvolt/\ele}.
Linearity of the detector response, assumed in this method, is validated in \cite{APTSSF_2023,opamp_2025}.

\paragraph{Clustering}
The data provided by the DUT, trigger, and tracking planes were processed using the Corryvreckan software framework \cite{corryvreckan}. 
Clusters were formed from adjacent pixels, applying a clusterization threshold of \SI{100}{e^-}. For analog sensors, only signals exceeding an offline-defined charge threshold of three time the noise RMS (cf. Section~\ref{sec:noise})
were considered. Signal amplitudes were converted to electrons using the calibration described earlier. The seed pixel was identified as the one with the highest charge in the cluster, and the cluster position was determined via a charge-weighted center-of-gravity method.
\paragraph{Track reconstruction}
The Corryvreckan framework was used for track reconstruction, and employs the General Broken Lines (GBL) algorithm \cite{Blobel:2006yi} to fit clusters on the reference planes, accounting for multiple scattering. Software alignment was carried out by iteratively shifting and rotating planes relative to a fixed reference to optimize tracking precision. To ensure unbiased results, the DUT was excluded from the fitting of tracks. The track position on the DUT plane was determined by interpolation.
The quality of the tracking was granted by the following selections: a single cluster was required on every tracking plane, the tracks were selected based on fit quality (${\chi}^2/n_\text{dof}<5$), only single-track events triggered by the APTS-SF were accepted, a region of interest (ROI) was defined to include only tracks passing through one of the three central pixels read out via oscilloscope. An association radius of \SI{25}{\um} was used to match DUT clusters to tracks, which demonstrated to provide unbiased efficiency estimation in Ref.~\cite{opamp_2025}. This value was optimized to ensure the efficiency overestimation is less than 0.05\%.
\paragraph{Detection efficiency analysis}
Detection efficiency was evaluated as the fraction of tracks crossing the ROI that had an associated cluster in one of the three oscilloscope pixels of the DUT. 
The above mentioned \SI{25}{\um} association radius is intended to reduce bias, ensuring the efficiency overestimation is less than 0.05\% \cite{opamp_2025}, given the tracking resolution of $\sigma_{track} = 2.7 \pm$ \SI{0.1}{\um}, obtained using a telescope optimizer simulation tool \cite{mager_telescope}. 
Baseline fluctuations were also analyzed to assess noise. The results are reported using thresholds above 
three times the maximum RMS noise.
\paragraph{Time resolution analysis}
Time residuals are computed as the difference between the APTS-OA DUT seed pixel time at 10\% amplitude and the LGAD reference at 40\%, where the LGAD resolution $\sigma_\textrm{LGAD}=$ \SI{23}{ps}~$\pm$~\SI{2}{ps} \cite{Carnesecchi_2023}. 
This constant-fraction method minimizes time walk and yields the lowest RMS, as demonstrated in \cite{opamp_2025}. 
The DUT time resolution is extracted by quadratically subtracting the LGAD resolution from the standard deviation of the time residuals:
\begin{equation} \label{eq:time_resolution}
    \sigma_{\textrm{t}} = \sqrt{\sigma^2_{\textrm{$\Delta$\textit{t}}} - \sigma^2_{\textrm{LGAD}}} \ ,
\end{equation}
assuming statistical independence.

\subsection{Results and discussion}
\paragraph{Noise}
\label{sec:noise}
The noise was characterized from a series of repeated measurements of a fixed baseline point acquired at constant bias, and the RMS of the corresponding distribution was used as the noise estimate.
\begin{figure}[!ht]
    \centering
    \begin{subfigure}[t]{.49\textwidth}
        \centering
        \includegraphics[width=\linewidth]{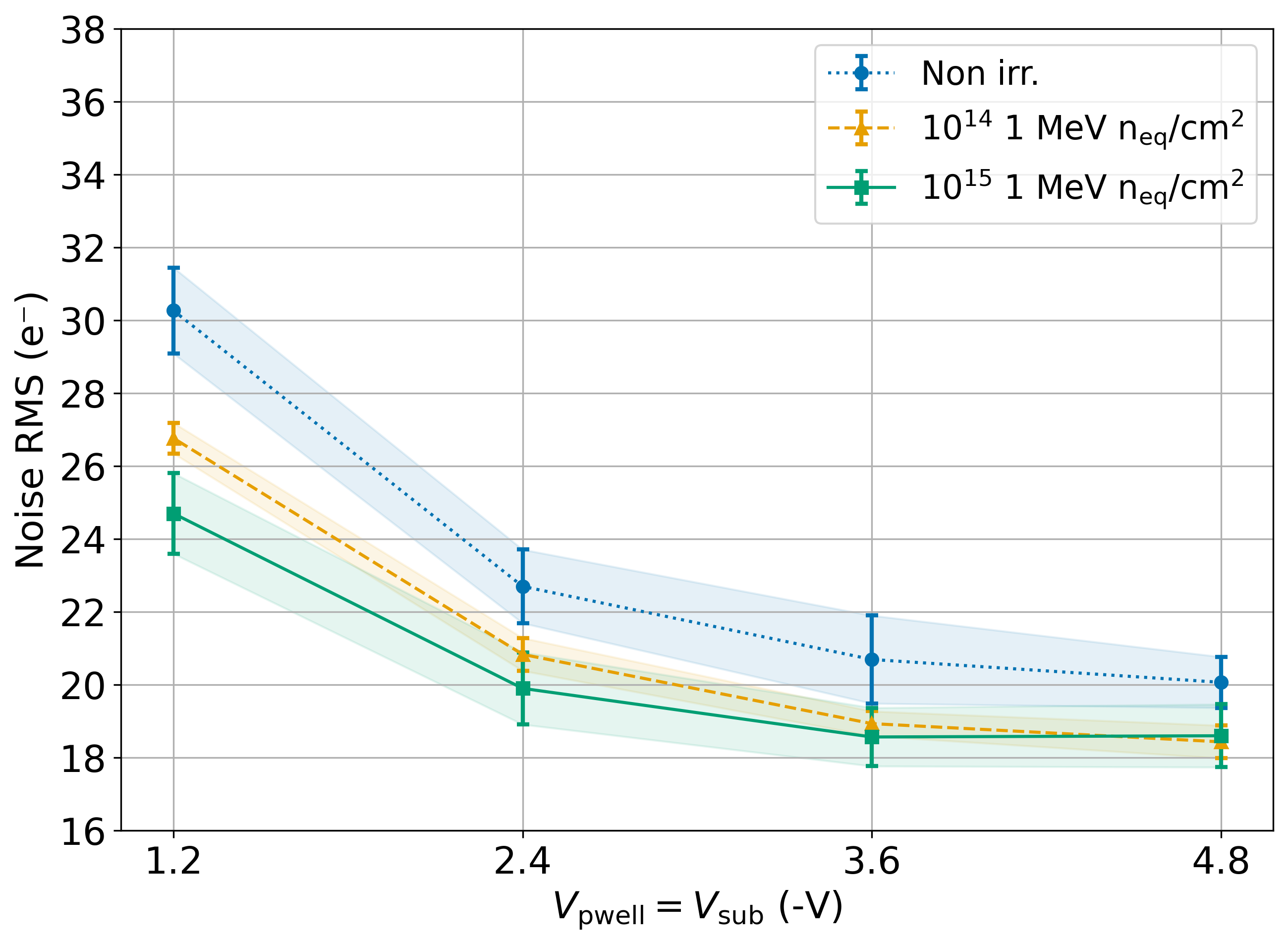}
        \caption{}
        \label{fig:noise_irr}
    \end{subfigure}
    \begin{subfigure}[t]{.49\textwidth} 
        \centering
        \includegraphics[width=\linewidth]{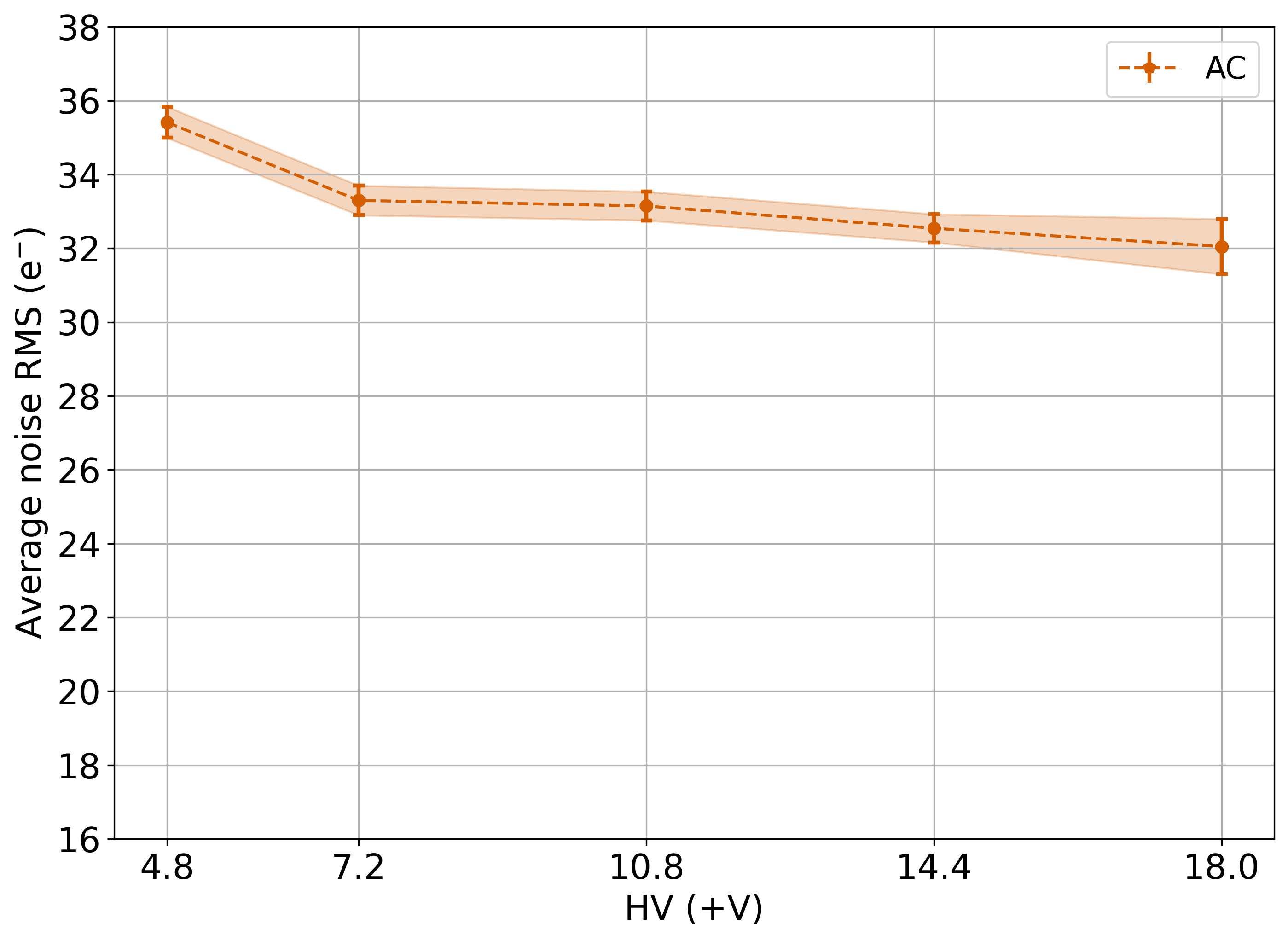}
        \caption{}
        \label{fig:noise_AC}
    \end{subfigure}
    \caption{The average noise RMS of the three central pixels read out using an oscilloscope versus the applied depletion voltage for DC-coupled irradiated (a) and AC-coupled sensors (b).}
    \label{fig:noise}
\end{figure}
Figure~\ref{fig:noise} presents the noise RMS as a function of the applied bias voltage. For the DC-coupled irradiated sensors, shown in Figure~\ref{fig:noise_irr}, the curves demonstrate a decrease in the total noise level with increasing irradiation damage. 
Two competing mechanisms influence the noise after irradiation.
On one side, the higher leakage current increases the intrinsic sensor noise. On the other hand, irradiation reduces the sensor capacitance. Because the front-end noise can be represented as a series voltage source~\cite{Rivetti}, the corresponding equivalent noise charge is proportional to the input capacitance. A lower capacitance therefore decreases the equivalent noise charge, partially compensating the intrinsic noise increase.
It is worth noting that the APTS-SF showed an opposite result for similar sensors, with increasing noise levels after irradiation~\cite{APTSSF_2023}. This was finally understood to be caused by the much lower sampling rate of 4 MS/s in the test setup used for the APTS-SF. The low sampling rate introduces a sampling error in the signal, which manifested ad an additional noise source in the measurement.
This effect is more significant for irradiated sensors, with a faster return to zero due to a larger reset current setting to accommodate the larger sensor leakage \cite{sanna2025poster, corentin2025thesis}.


The AC-coupled sensor (Figure~\ref{fig:noise_AC}) exhibits a higher RMS noise, mainly due to 
the coupling capacitor and the additional parasitic capacitive load it introduces, leading to a lower voltage excursion at the circuit input for the same signal charge.
Both AC-coupled and irradiated sensors exhibit a reduction in noise with increasing substrate bias voltage, consistent with the trends reported in previous studies~\cite{opamp_2025}. 

To ensure robust signal discrimination, a threshold of  more than three times the RMS noise distribution has been applied. Timing analysis was studied only at a threshold of \SI{100}{\ele}.

\begin{figure}[!ht]
    \centering
    \begin{subfigure}[t]{.49\textwidth}
        \centering
        \includegraphics[width=\linewidth]{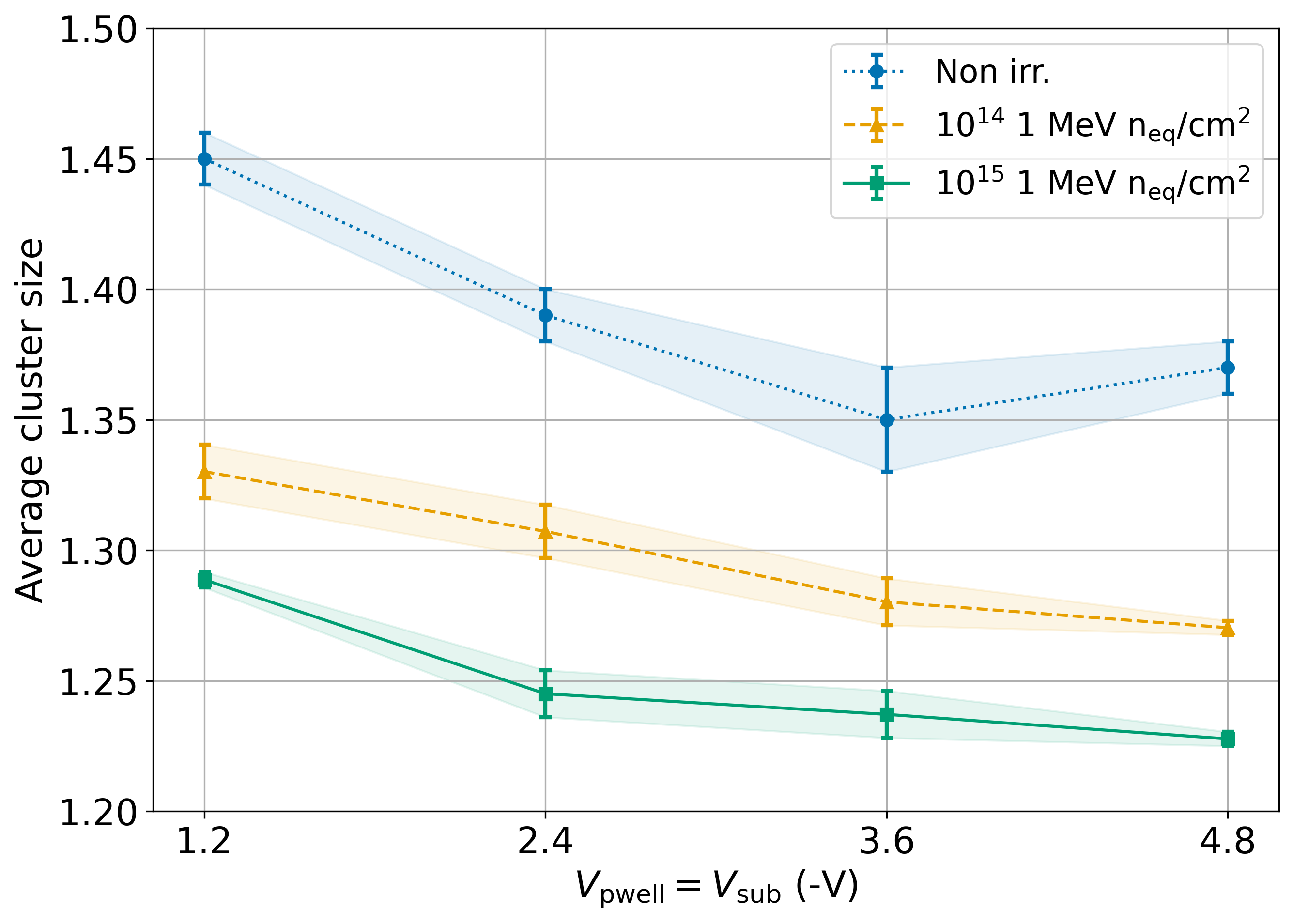}
        \caption{}
        \label{fig:cs_vdepl_irr}
    \end{subfigure}
    \begin{subfigure}[t]{.49\textwidth}
        \centering
        \includegraphics[width=\linewidth]{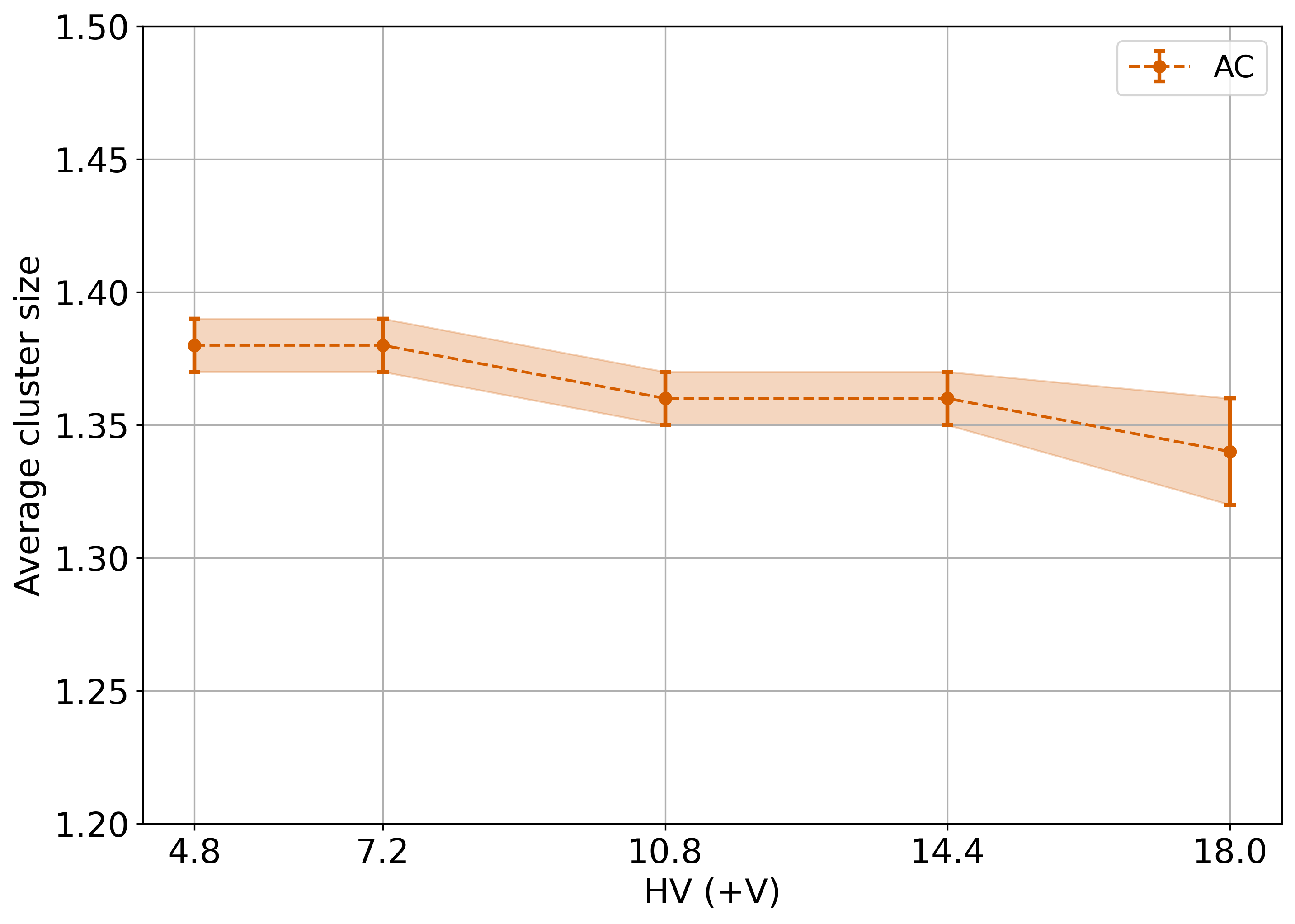}
        \caption{}
        \label{fig:cs_vdepl_AC}
    \end{subfigure}
    \caption{The average estimated cluster size versus the applied reverse bias voltage for DC-coupled irradiated sensors (a) and AC-coupled sensor (b).}
    \label{fig:cs_vdepl}
\end{figure}

\paragraph{Cluster size}
Figure~\ref{fig:cs_vdepl} shows the average cluster size as a function of the applied substrate voltage and HV for DC-coupled and AC-coupled sensors respectively. 
A clear decreasing trend is observed with increasing bias, which is expected: as the reverse bias increases, the epitaxial layer of the sensor becomes more depleted. This results in reduced charge sharing between pixels and, consequently, smaller cluster sizes.\cite{opamp_2025}
The AC-coupled sensor (Figure~\ref{fig:cs_vdepl_AC}) follows a trend consistent with the DC-coupled case. The irradiated sensors (Figure~\ref{fig:cs_vdepl_irr}) exhibit a lower average cluster size, attributed to charge capturing by trapping centers introduced by irradiation.

\paragraph{Charge collection}
The cluster charge is defined as the total charge collected by all pixels belonging to a reconstructed cluster.
\begin{figure}[!ht]
    \centering
    \begin{subfigure}[t]{.49\textwidth}
        \centering
        \includegraphics[width=\linewidth]{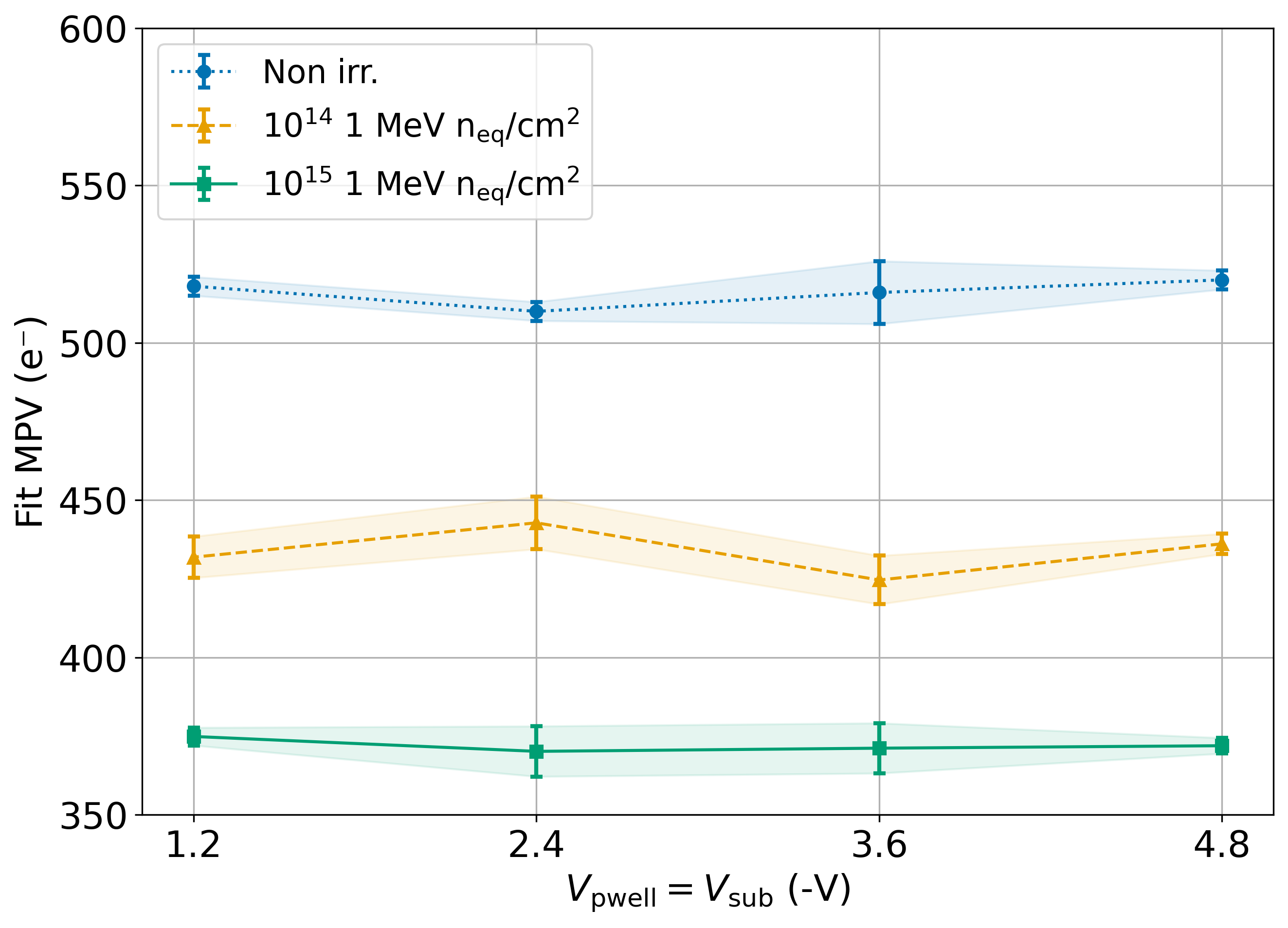}
        \caption{}
        \label{fig:landau_vdepl_irr}
    \end{subfigure}
    \begin{subfigure}[t]{.49\textwidth} 
        \centering
        \includegraphics[width=\linewidth]{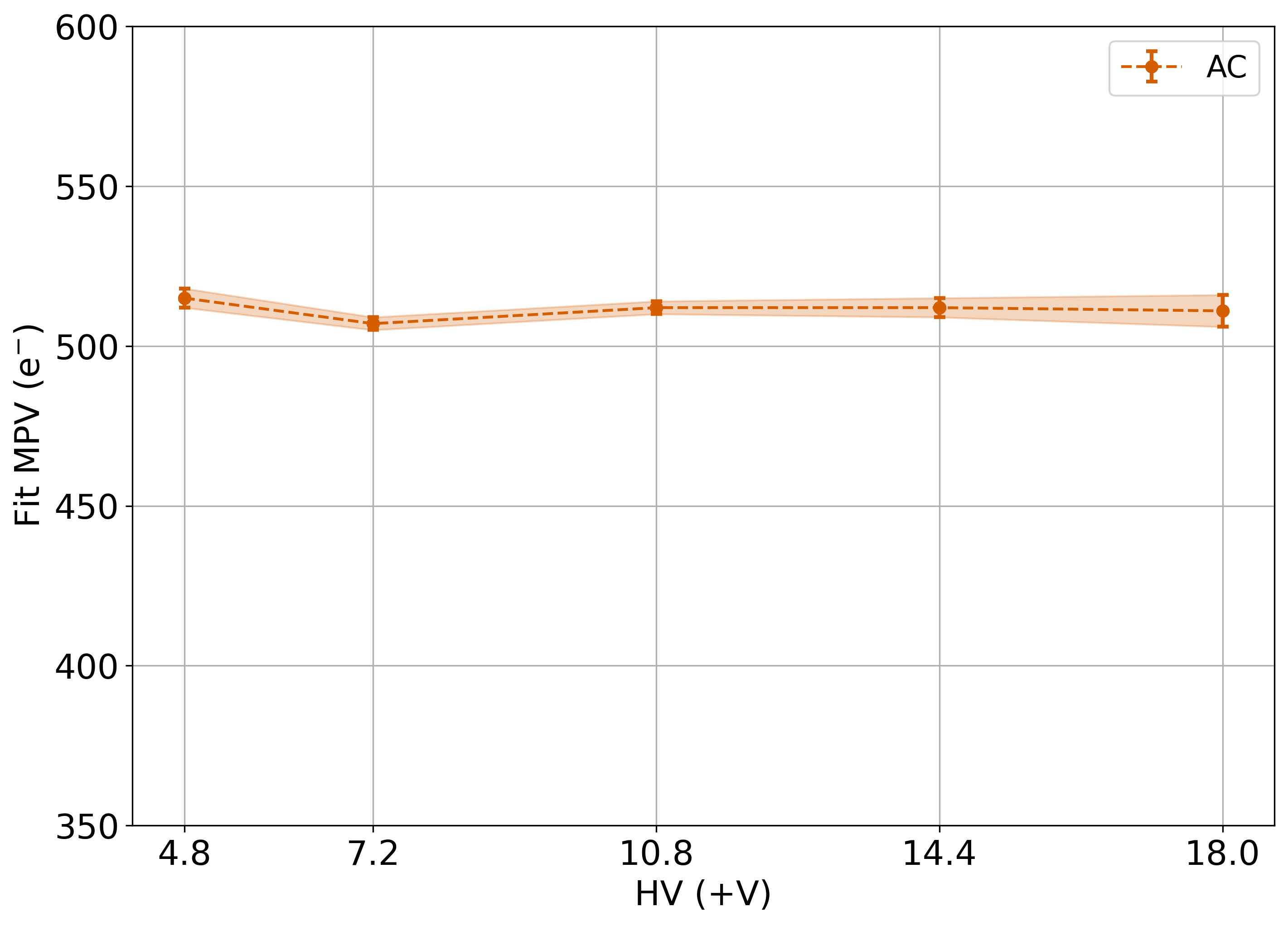}
        \caption{}
        \label{fig:landau_vdepl_AC}
    \end{subfigure}
    \caption{MPV calculated from Landau-Gauss fit of the charge collection distributions versus the applied reverse bias voltage for DC-coupled irradiated sensors (a) and AC-coupled sensors (b). The errors are statistical only.}
    \label{fig:landau_vdepl}
\end{figure}
Figure~\ref{fig:landau_vdepl} shows the most probable value (MPV) of the cluster charge collection distributions for the DC-coupled irradiated (Fig.~\ref{fig:landau_vdepl_irr}) and AC-coupled sensor (Fig.~\ref{fig:landau_vdepl_AC}). The cluster charge distributions 
can be found in \ref{apx:cluster_charge}. In both AC and DC-coupled sensors, the MPV remains largely independent of the applied bias voltage. This is expected since the input capacitance suppression is compensated while applying the charge calibration. In addition, the efficiency at this threshold (\SI{100}{e^-}) is very close to 100\%, therefore all the charge is collected for any bias voltage.

While the AC-coupled sensor exhibits an MPV consistent with that of the non-irradiated DC-coupled sensor, the irradiated devices show a significant reduction in the collected charge. This loss is attributed to the increased number of trapping centers induced by radiation damage, consistent with the behavior previously observed in the cluster size distributions.

\begin{figure}[!ht]
    \centering
    \begin{subfigure}[t]{.45\textwidth}
        \centering
        \includegraphics[width=\linewidth]{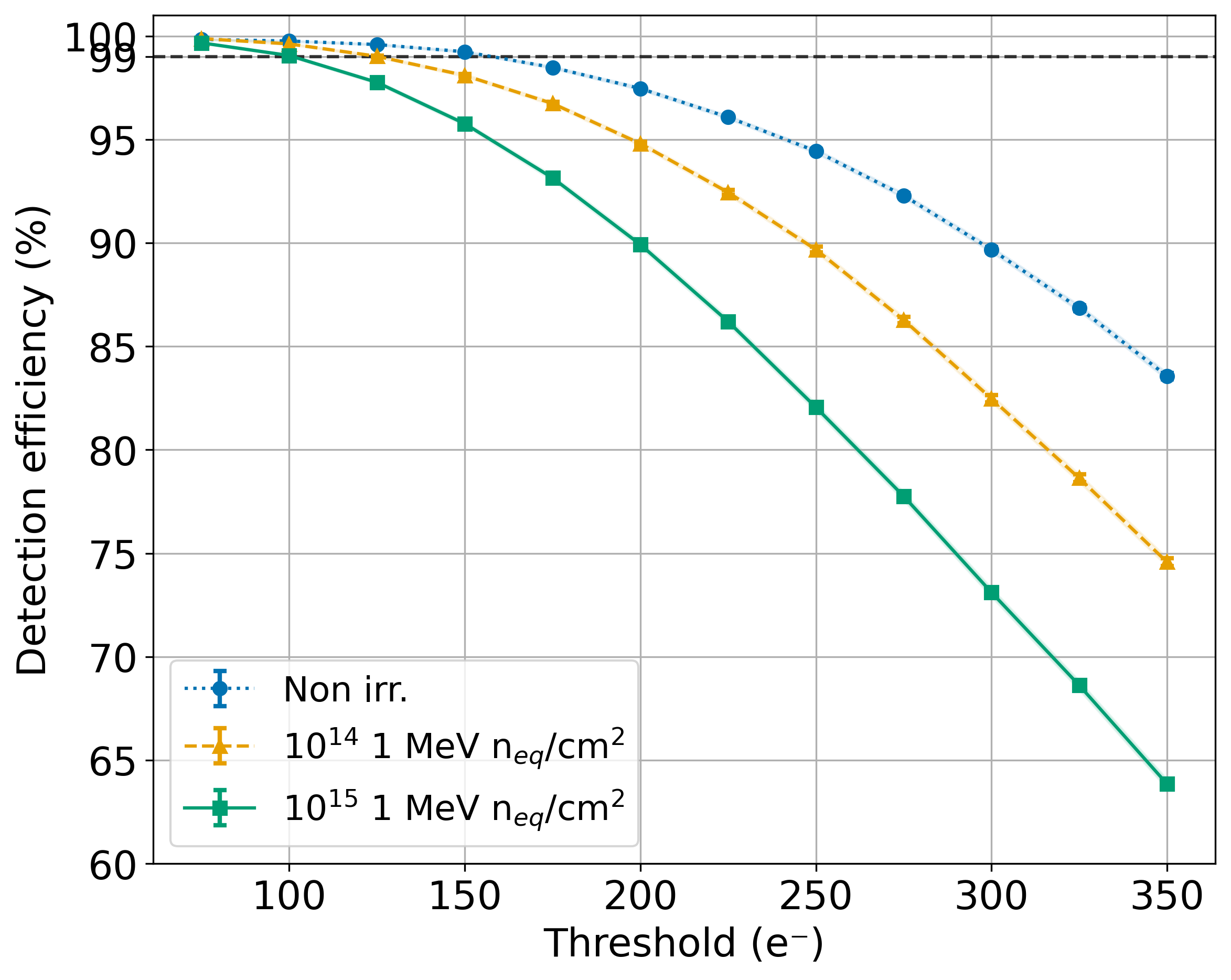}
        \caption{}
        \label{fig:efficiency_irr}
    \end{subfigure}
    \begin{subfigure}[t]{.51\textwidth} 
        \centering
        \includegraphics[width=\linewidth]{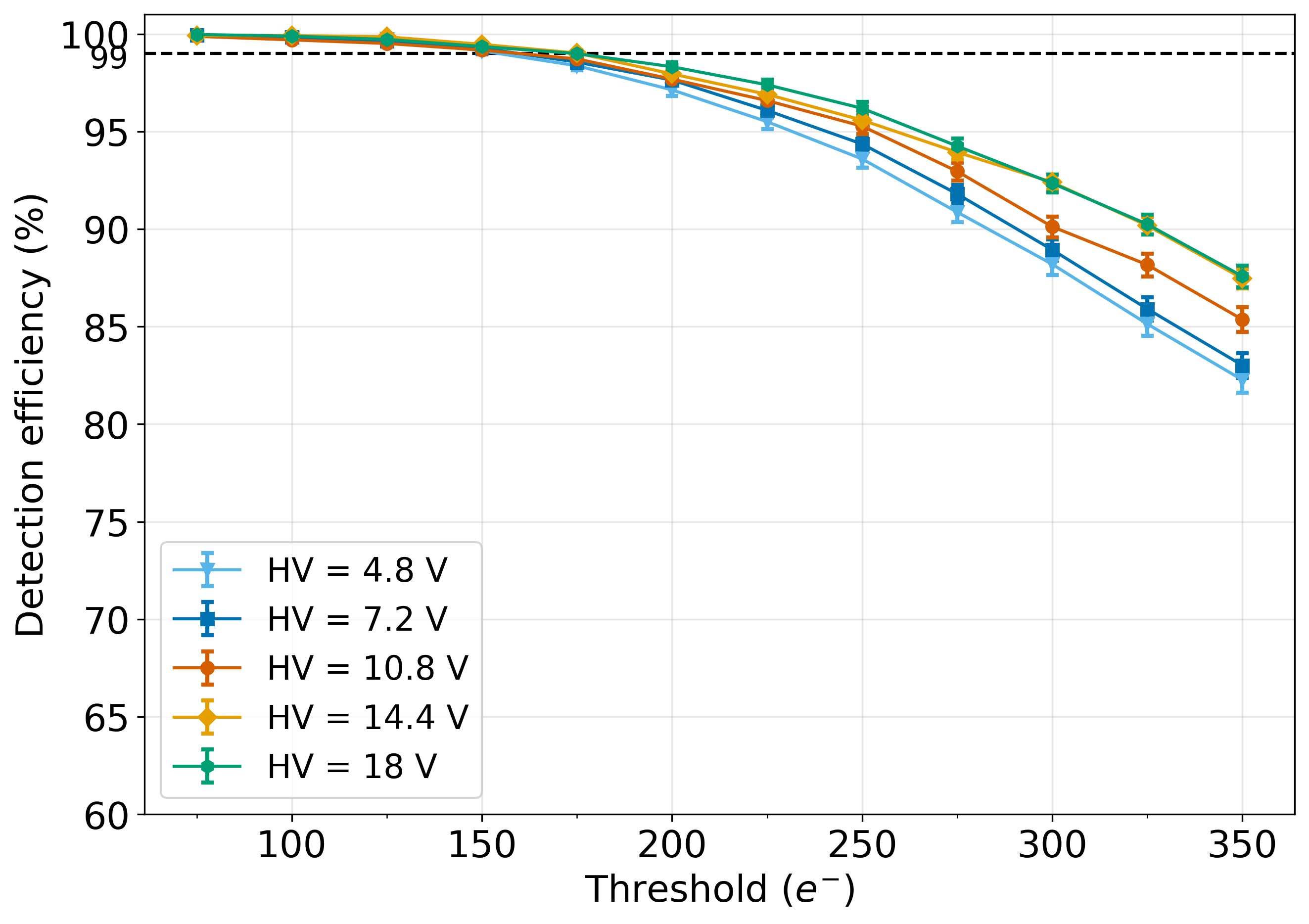}
        \caption{}
        \label{fig:efficiency_AC}
    \end{subfigure}
    \caption{Detection efficiency as a function of the applied threshold of DC-coupled irradiated sensors at $V_{\rm{sub}}=\SI{-4.8}{\volt}$ (a) and AC-coupled sensor at different diode voltages (b). The horizontal dashed line indicates 99\% detection efficiency.}
    \label{fig:efficiency}
\end{figure}

\paragraph{Detection efficiency}
Figure~\ref{fig:efficiency} presents the detection efficiency for the irradiated sensors (Figure~\ref{fig:efficiency_irr}) and AC-coupled sensor (Figure~\ref{fig:efficiency_AC}) as a function of the threshold, for different irradiation levels and reverse bias voltages respectively. 
As expected, a clear decrease in detection efficiency is observed with increasing threshold. Both design variants—non-irradiated DC-coupled and AC-coupled sensors—achieve a detection efficiency greater than 99\% up to a threshold of \SI{150}{\ele}, in agreement with the results reported in Ref.~\cite{opamp_2025}, thus offering a robust operational margin. 

\begin{figure}[!ht]
    \centering
    \begin{subfigure}[t]{.45\textwidth}
        \centering
        \includegraphics[width=\linewidth]{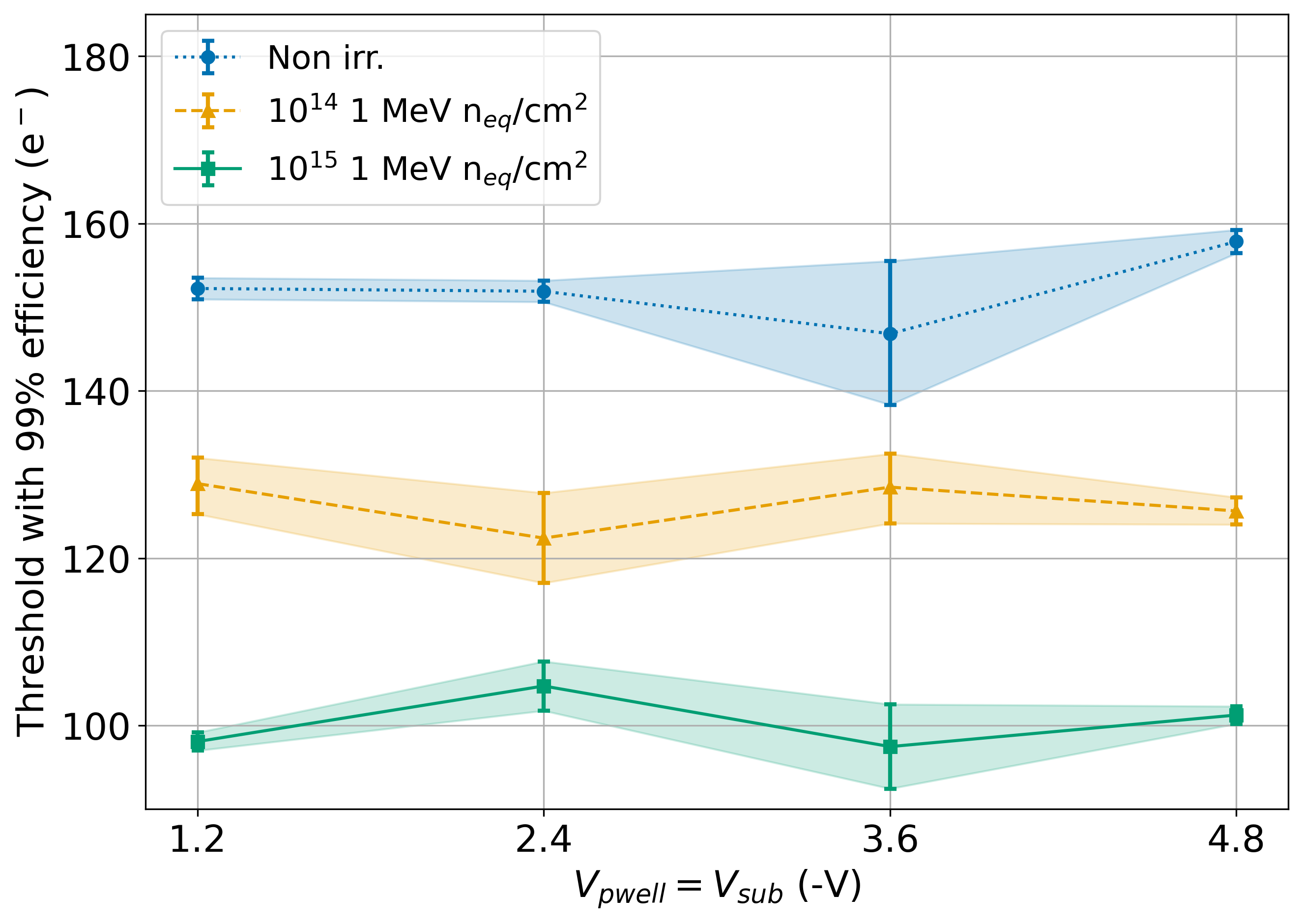}
        \caption{}
        \label{fig:efficiency_99p_irr}
    \end{subfigure}
    \begin{subfigure}[t]{.45\textwidth} 
        \centering
        \includegraphics[width=\linewidth]{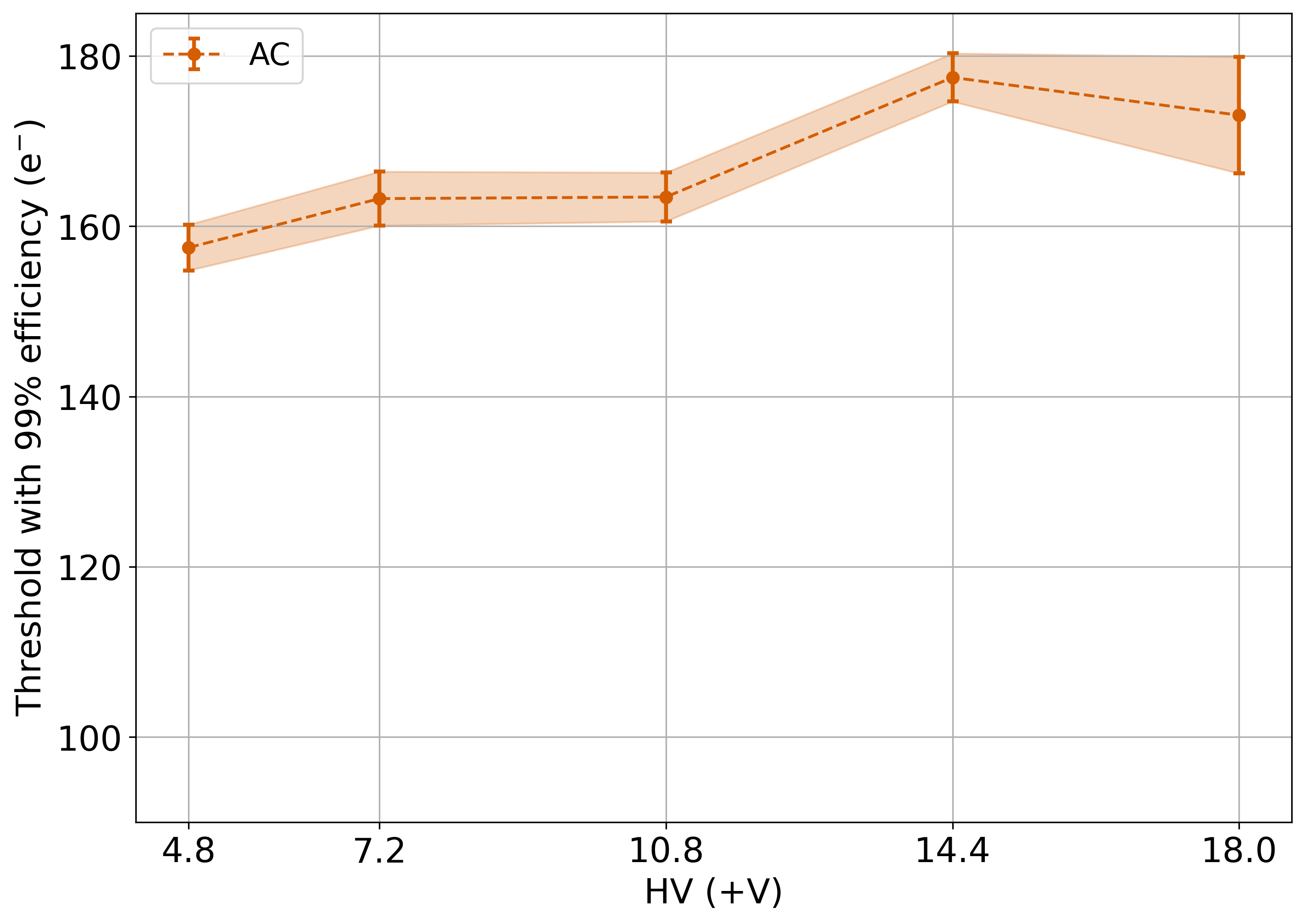}
        \caption{}
        \label{fig:efficiency_99p_AC}
     \end{subfigure}
     \caption{Maximum threshold at which the sensor achieves 99\% detection efficiency as a function of the applied reverse bias voltage for DC-coupled irradiated sensors (a) and AC-coupled sensors (b).}
     \label{fig:efficiency_99p}
\end{figure}

\begin{figure}[!ht]
    \centering
    \begin{subfigure}[t]{.49\textwidth}
        \centering
        \includegraphics[width=\linewidth]{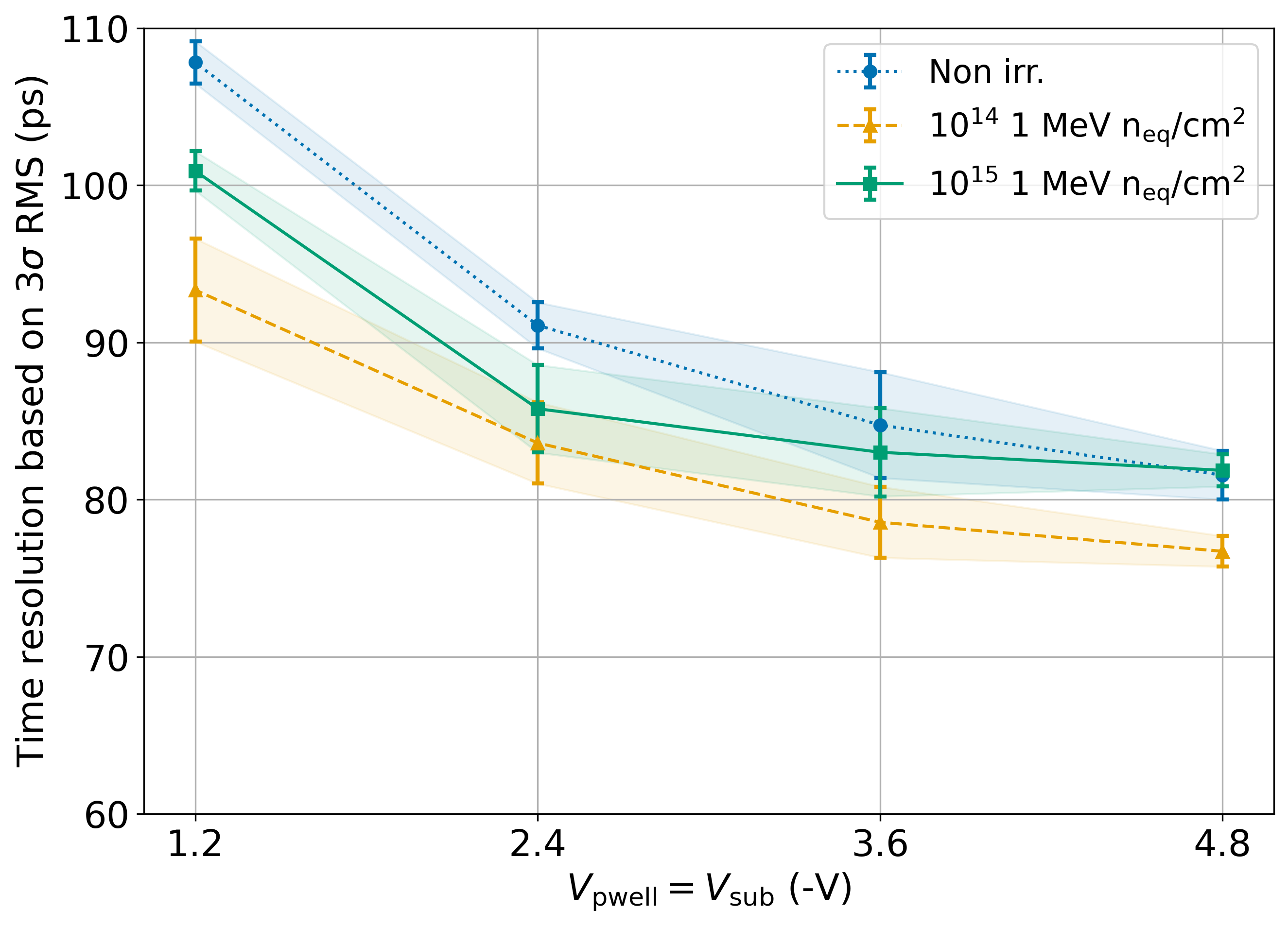}
        \caption{}
        \label{fig:3sigma_rms_vsub_nocorr_irr}
    \end{subfigure}
    \begin{subfigure}[t]{.49\textwidth} 
        \centering
        \includegraphics[width=\linewidth]{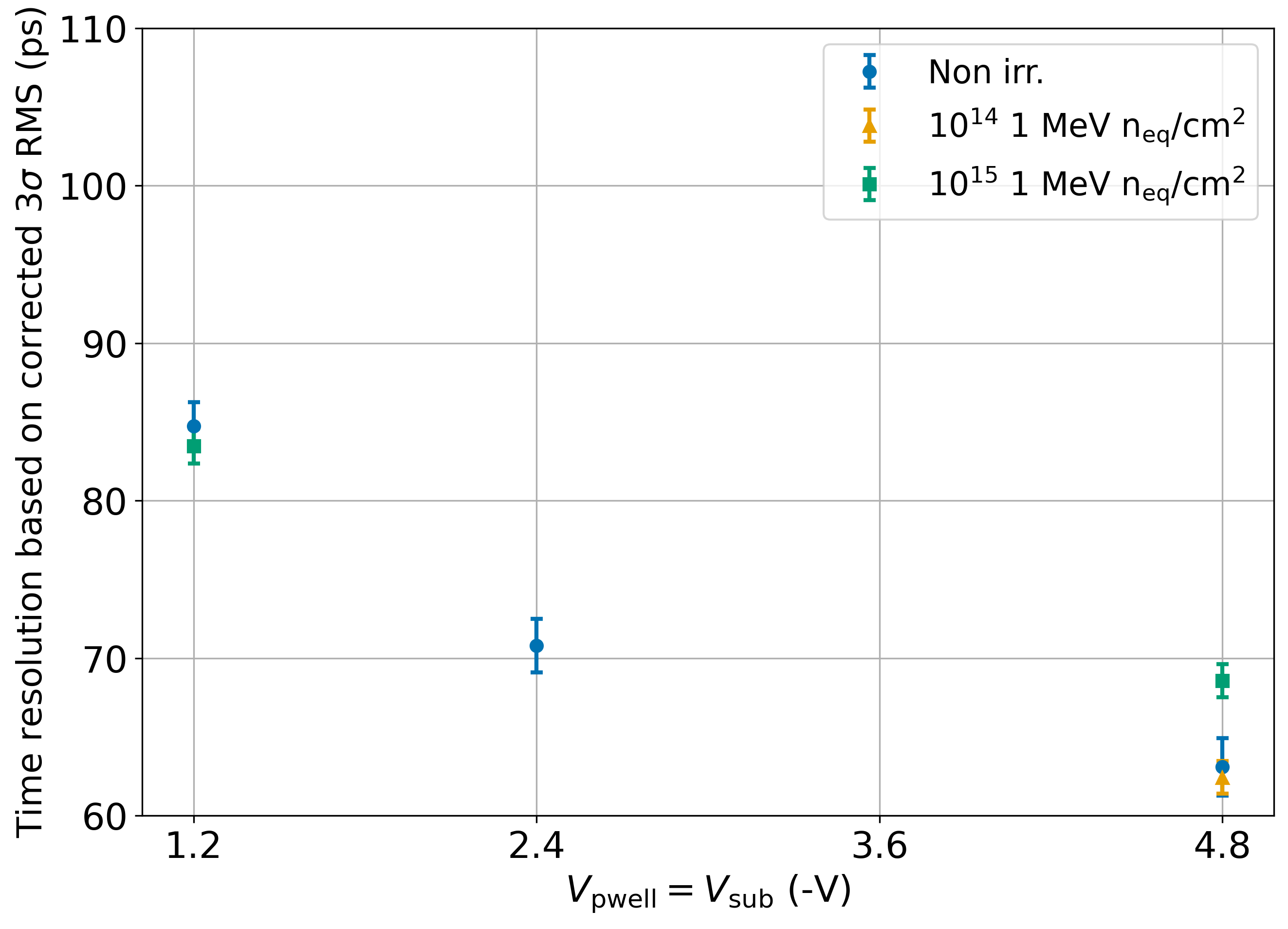}
        \caption{}
        \label{fig:3sigma_rms_vsub_corr_irr}
    \end{subfigure}
    \caption{The time residual width versus the applied reverse bias voltage of DC-coupled irradiated sensors before (a) and after (b) a time walk-like correction. The time resolution is determined by quadratically subtracting the time resolution of the LGAD used as a time reference from the 3$\sigma$ standard deviation of the time residuals.}
    \label{fig:3sigma_rms_vsub_irr}
\end{figure}

A more detailed study of NIEL effects on charge collection efficiency is reported in Figure~\ref{fig:efficiency_99p_irr}, showing the maximum threshold at which 99\% detection efficiency is reached as a function of reverse bias voltage. 
As the irradiation level increases, an efficiency of 99\% can only be maintained at lower threshold values, reducing, as expected, the operational margin.
These results further support the interpretation of more effective charge recombination at the pixel periphery, as suggested by the reduced average cluster size (Figure~\ref{fig:cs_vdepl_irr}), as well as the overall charge-collection degradation (Figure~\ref{fig:landau_vdepl_irr}) with increasing NIEL. The corresponding impact on time resolution will be shown in Figure~\ref{fig:3sigma_rms_vsub_irr}.


\begin{figure}[!ht]
    \centering
    \begin{subfigure}[t]{.49\textwidth}
        \centering
        \includegraphics[width=\linewidth]{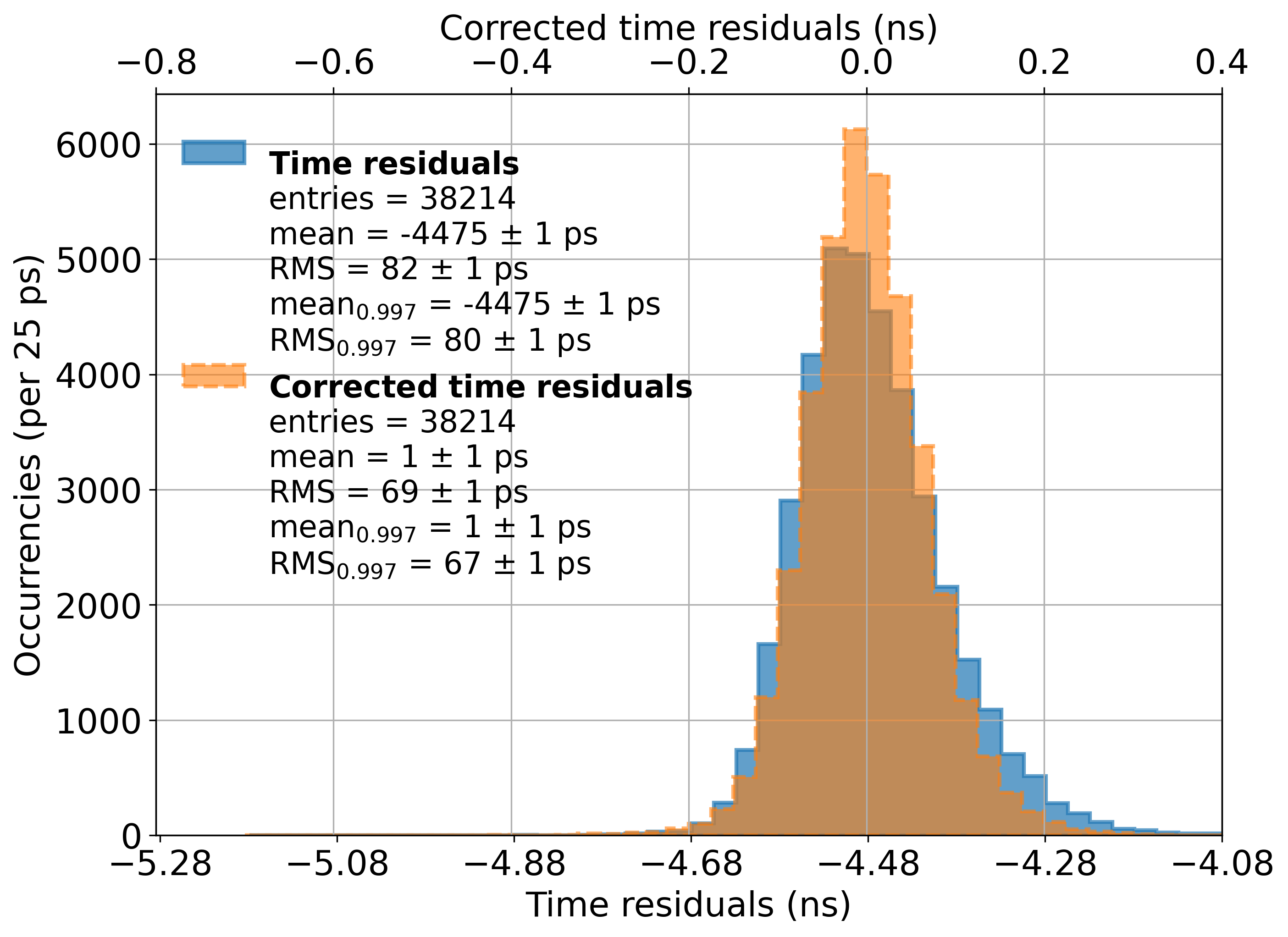}
        \caption{}
        \label{fig:time_res_Pi1}
    \end{subfigure}
    \begin{subfigure}[t]{.49\textwidth} 
        \centering
        \includegraphics[width=\linewidth]{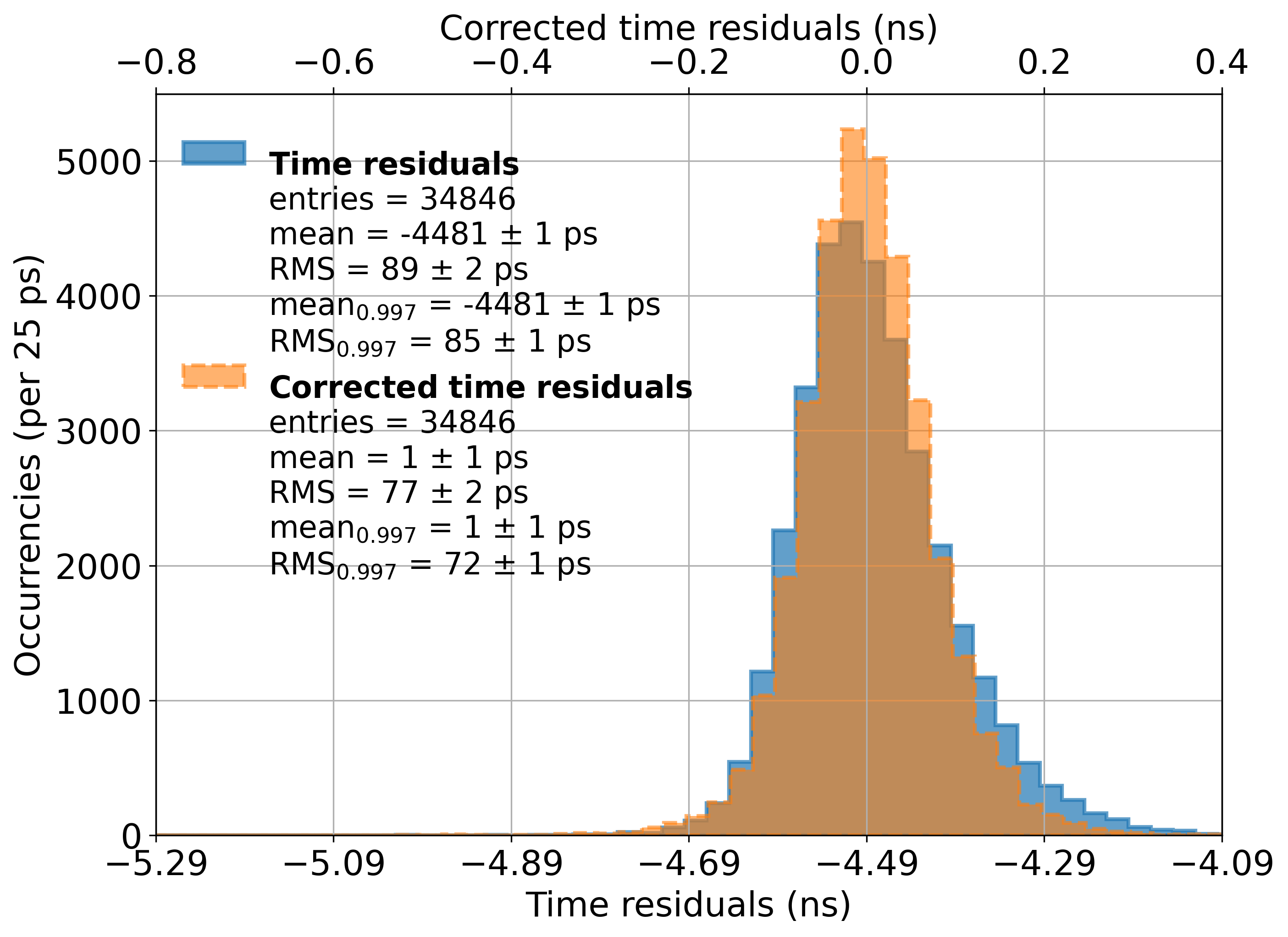}
        \caption{}
        \label{fig:time_res_Pi3}
    \end{subfigure}
    \begin{subfigure}[t]{.49\textwidth} 
        \centering
        \includegraphics[width=\linewidth]{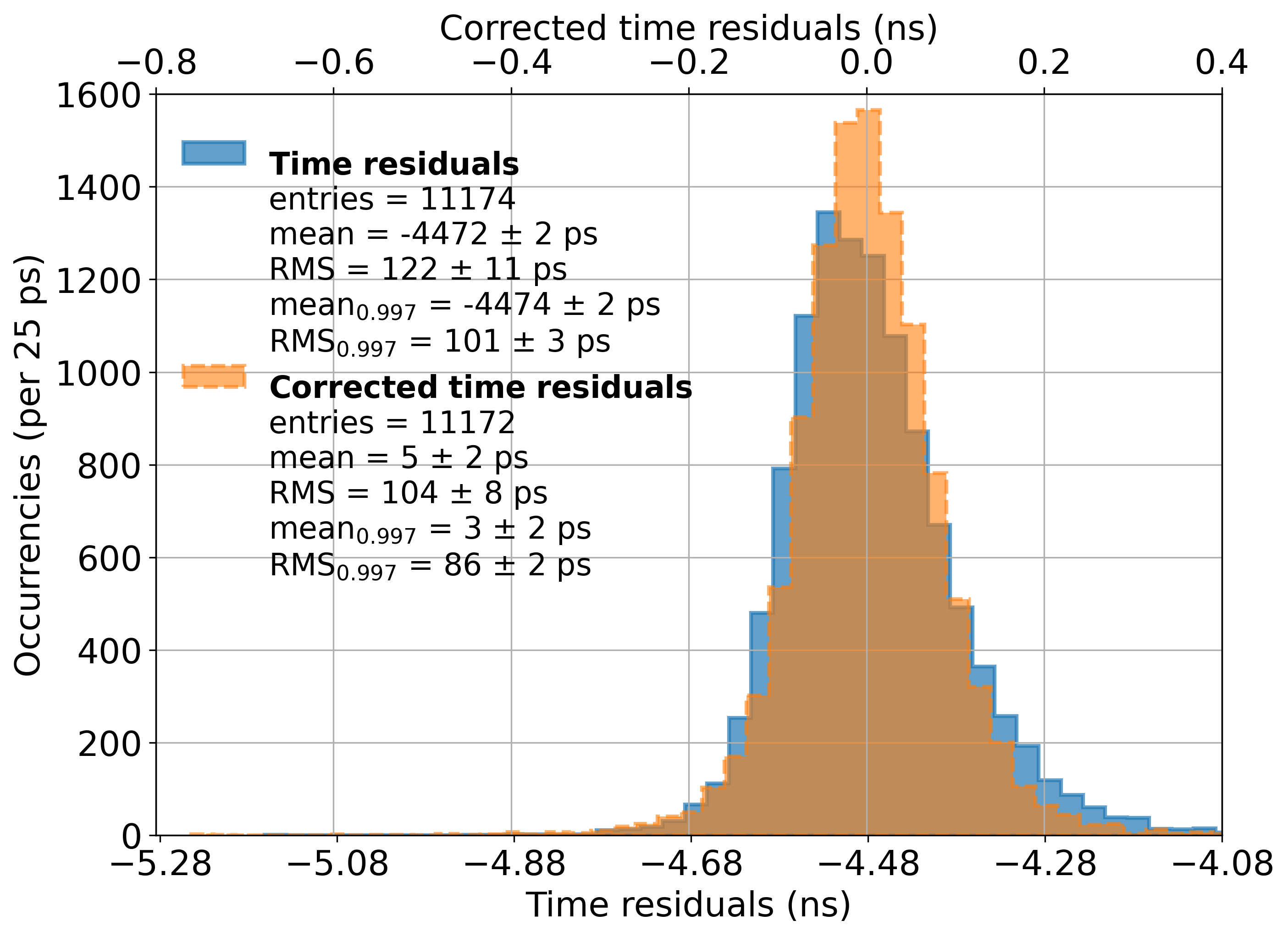}
        \caption{}
        \label{fig:time_res_AC}
    \end{subfigure}
    \caption{Time residuals of DC-coupled sensors irradiated to \SI{d14}{\NIEL} (a), \SI{d15}{\NIEL} (b) at $V_\text{sub}=\SI{-4.8}{\volt}$ and AC-coupled sensor at $V_{\rm{HV}}=\SI{4.8}{\volt}$ (c). The bottom horizontal axis corresponds to the time residual distribution before correction shown in blue. The top horizontal axis corresponds to the time residual distribution after a time walk-like correction shown in orange.}
    \label{fig:time_res}
\end{figure}

\paragraph{Timing}
Figure~\ref{fig:time_res} shows the time residuals of the DC-coupled irradiated sensors and AC-coupled sensor in blue. 
As previously reported \cite{opamp_2025}, the measured time reference signals are delayed relative to those from the DUT, resulting in negative time residuals (see Eq.~\ref{eq:time_resolution}). Despite efforts to minimize the influence of signal amplitude on timing by using constant fraction discrimination, the presence of a longer tail in the positive direction of the time residual distribution indicates residual time walk. 
This is further supported by a correlation between time residuals and the DUT seed signal amplitude, already observed in \cite{opamp_2025}.
%

Time walk persists because constant fraction discrimination only eliminates it when the signal shape is consistent, and in this case, the signal shape varies with both the amount of collected charge and the position of ionization within the pixel. Therefore, an additional correction based on cluster size is applied to minimize the contribution of time walk to the time residuals, as proposed in~\cite{Braach_2023}), classifying data into single- and multi-pixel clusters. For each group, a quadratic interpolation of mean time residuals vs. signal amplitude is performed. To avoid biasing the results, the interpolation is performed on half of the data and subtracted from the other half, and vice versa. The two corrected halves are finally merged for the analysis. Two metrics are used to evaluate the residual time spread: (1) the overall RMS, and (2) the RMS of the central 99.7\% (hereafter called $3\sigma$ RMS) of the distribution.

The time residual distributions for the irradiated sensors, shown in Figure~\ref{fig:time_res_Pi1} and ~\ref{fig:time_res_Pi3}, exhibit an asymmetric shape with a $3\sigma$ RMS. 
The difference becomes more pronounced at lower reverse bias voltages, as illustrated in Figure~\ref{fig:3sigma_rms_vsub_nocorr_irr}. 
This behavior is attributed to a higher probability of charge recombination for hits farther from the collection electrode, arising from radiation-induced trapping centers, as confirmed by the cluster size (Fig.\ref{fig:cs_vdepl}). As a result, the tail at high time residuals is more suppressed comparing to the non-irradiated sensor \cite{opamp_2025}, leading to a reduced overall RMS.

When applying residual-time-walk correction, the $3\sigma$ RMS values become compatible between irradiated and non-irradiated sensors at low substrate bias (Figure~\ref{fig:3sigma_rms_vsub_corr_irr}).
The impact of radiation on the time resolution is driven by two competing mechanisms.
First, radiation damage degrades the time resolution, mainly due to signal loss caused by reduced electric field and carrier trapping, as seen in the sensor irradiated to \SI{d15}{\NIEL}.
Second, a partial compensating effect arises because radiation preferentially suppresses the slow component of the signal, most evident in the sensor irradiated to \SI{d14}{\NIEL}.

\begin{figure}[!ht]
    \centering
    \begin{subfigure}[t]{.49\textwidth}
        \centering
        \includegraphics[width=\linewidth]{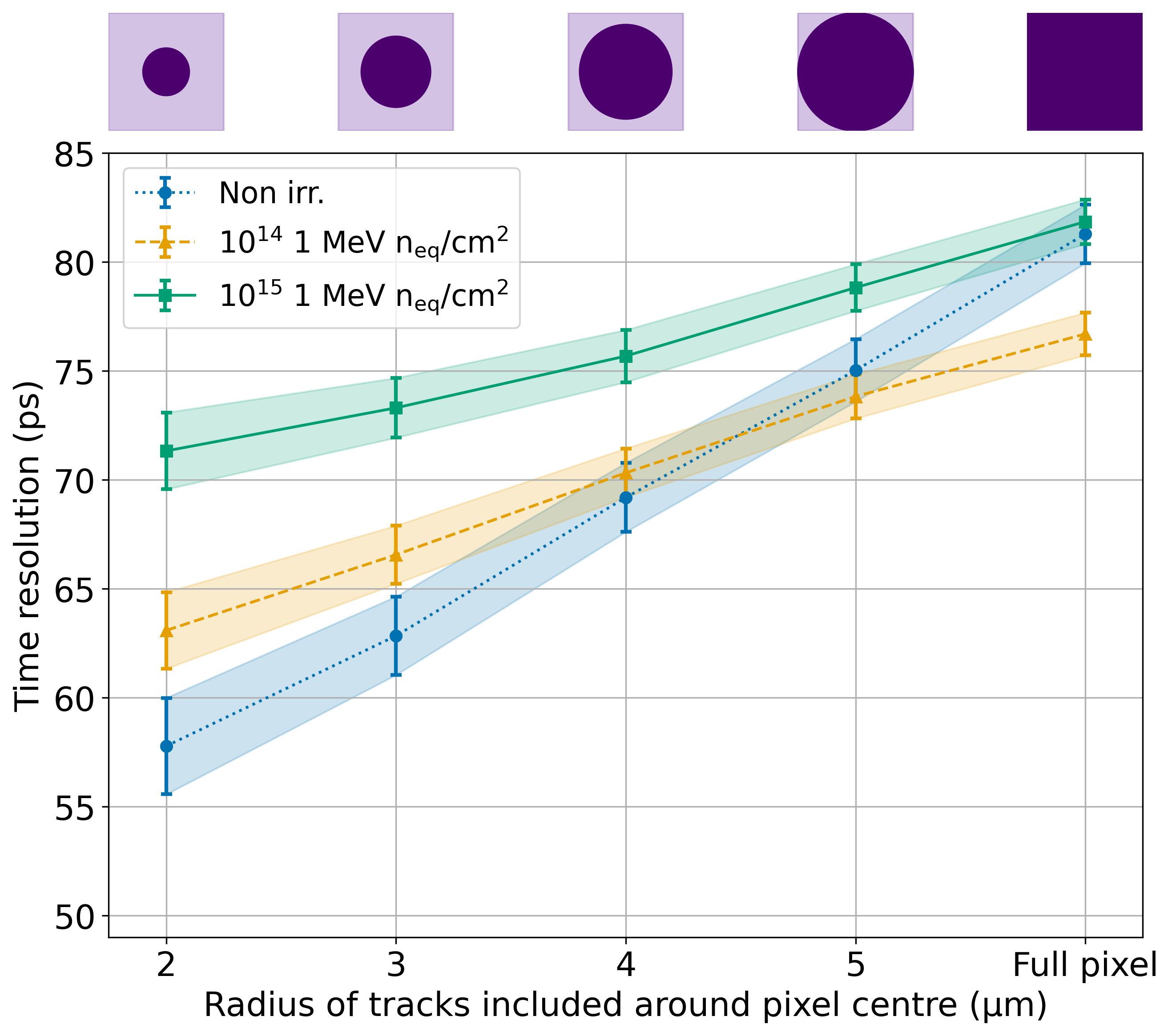}
        \caption{}
        \label{fig:3sigma_rms_cuts_nocorr_DC}
    \end{subfigure}
    \begin{subfigure}[t]{.49\textwidth} 
        \centering
        \includegraphics[width=\linewidth]{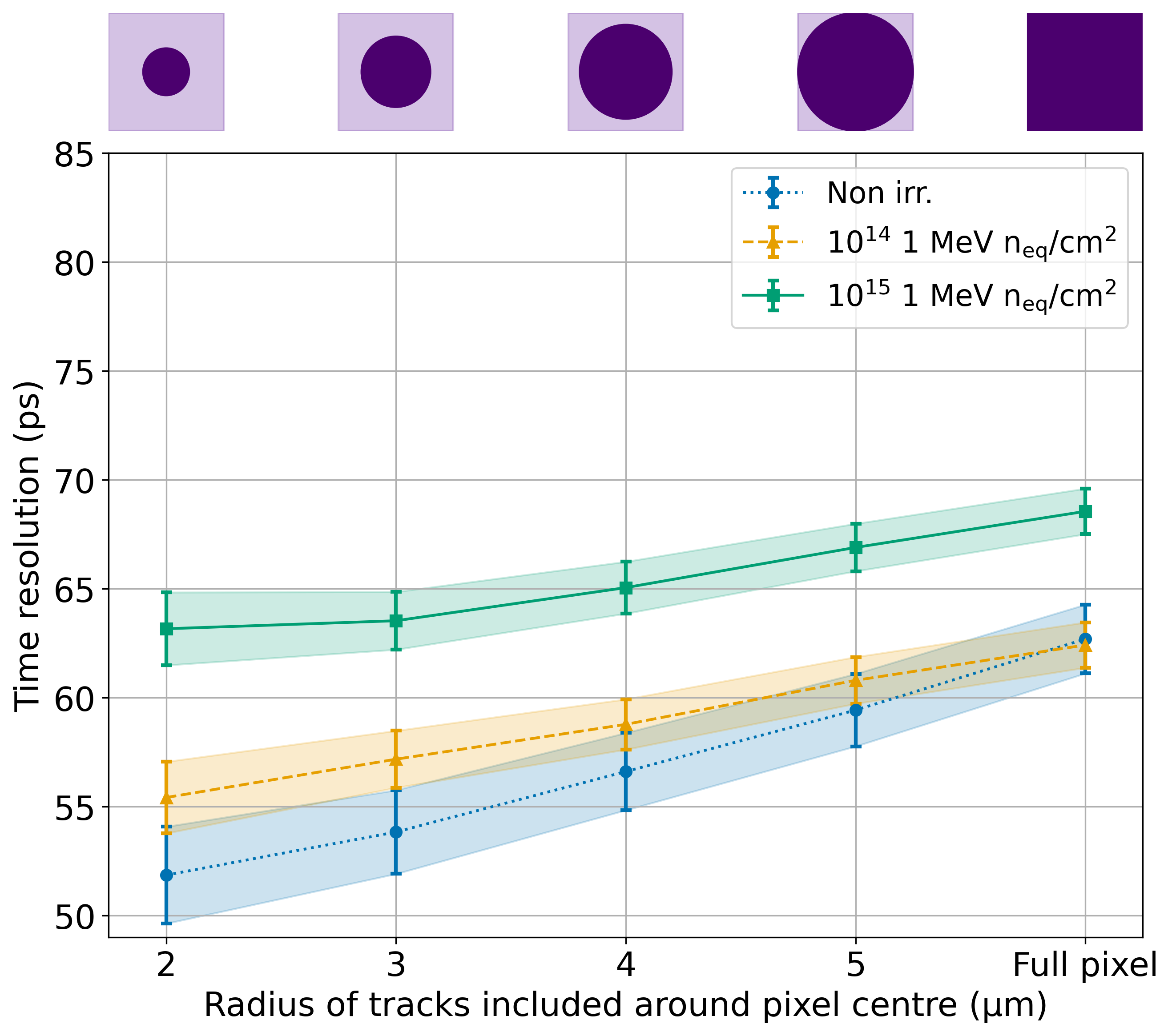}
        \caption{}
        \label{fig:3sigma_rms_cuts_corr_DC}
    \end{subfigure}
    \caption{Time resolution for different in-pixel cuts of DC-coupled irradiated sensors at $V_\text{sub}=\SI{-4.8}{\volt}$ before (a) and after (b) a time walk-like correction. The time resolution is determined by quadratically subtracting the time resolution of the LGAD used as a time reference from the 3$\sigma$ standard deviation of the time residuals.}
    \label{fig:3sigma_rms_vsub_irr_radii}
\end{figure}

To further investigate this effect, a radial selection within the pixel was performed using tracking-based spatial information. Figure~\ref{fig:3sigma_rms_vsub_irr_radii} shows that, at small distances from the collection electrode, the non-irradiated sensor demonstrates better time resolution. However, when including signals from the pixel periphery, a rapid degradation in timing performance is observed for the non-irradiated sensor. This increase is significantly milder in irradiated devices and is attributed to charge recombination reducing contributions from slow-drifting carriers near the sensor edges, as previously discussed.

A residual-time-walk correction proves to be an effective method for mitigating these peripheral effects.

In summary, under optimal operating conditions with full depletion at a substrate bias of \SI{-4.8}{\volt}, the sensor irradiated to \SI{d14}{\NIEL} achieves a time resolution of \SI[separate-uncertainty = true]{62.4\pm1.0}{\pico\second}, closely matching the \SI[separate-uncertainty = true]{62.7\pm1.6}{\pico\second} measured for the non-irradiated device. A performance degradation of approximately 10\% is observed for sensor irradiated to \SI{d15}{\NIEL}, for which a time resolution of \SI[separate-uncertainty = true]{68.6\pm1.0}{\pico\second} is measured.

\begin{figure}[!ht]
    \centering
    \includegraphics[width=.7\linewidth]{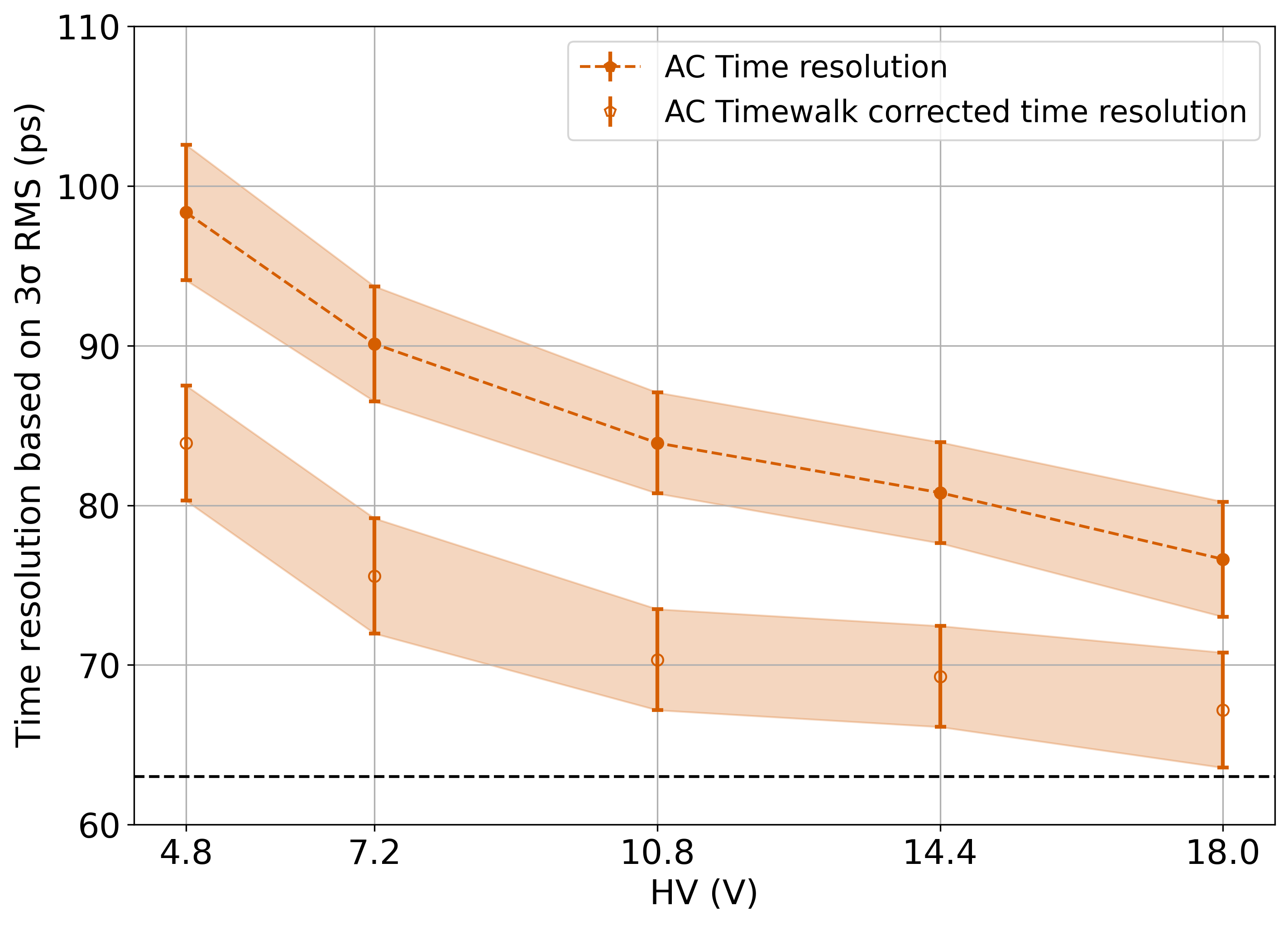}
    \caption{Time resolution versus the applied reverse bias voltage before (a) and after (b) a time walk-like correction. The errors are statistical only. The dashed line at $\SI{63}{\ps}$ represents the time resolution reached using a DC-coupled non irradiated sensor.}
    \label{fig:AC_3sigma_rms_vsub_AC}
\end{figure}

Regarding the results for the time resolution of the AC-coupled sensor, the effect of the increased electric field at the collection electrode was investigated by varying the high voltage from 4.8 to \SI{18}{\volt}.
As in the previous case, the time-walk correction is applied to the time residual distribution (Fig.~\ref{fig:time_res_AC}) to reduce its asymmetric shape.
The effect of the time-walk correction is shown in Figure~\ref{fig:AC_3sigma_rms_vsub_AC}. The overall time resolution improves consistently at all voltage values.
The $3\sigma$ RMS of the time distribution confirms this trend. It shows a progressively better timing performance with increasing bias, and reaching the value of \SI[separate-uncertainty = true]{67.2\pm3.6}{\pico\second} at \SI{18}{V}. At that bias, the time resolution becomes comparable to that of the DC-coupled sensor biased at \SI{-4.8}{\volt}~\cite{opamp_2025} within twice the uncertainty.


\begin{figure}[!ht]
    \centering
    \begin{subfigure}[t]{.49\textwidth}
        \centering
        \includegraphics[width=\linewidth]{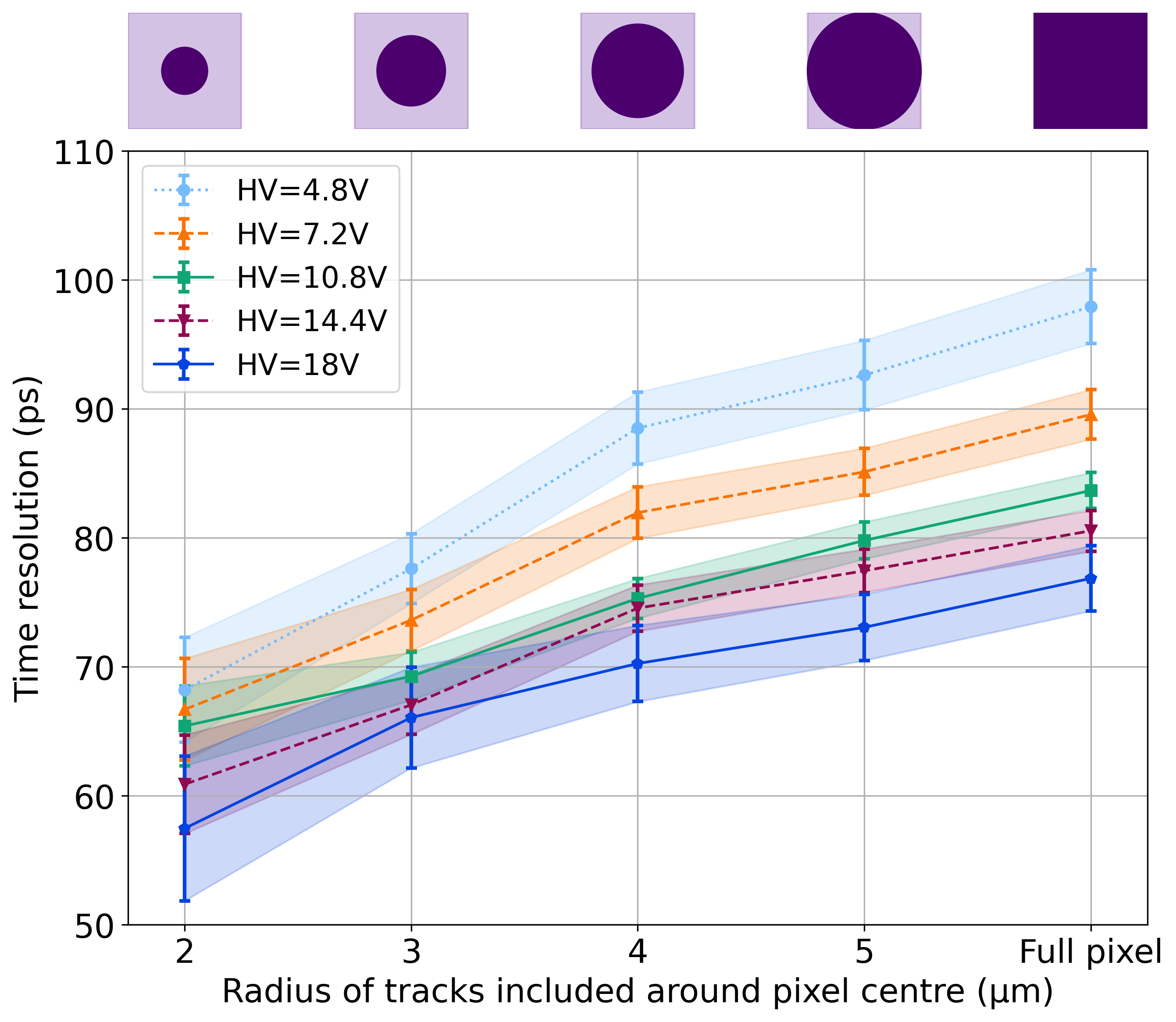}
        \caption{}
        \label{fig:3sigma_rms_cuts_nocorr_AC_merged_HV}
    \end{subfigure}
    \begin{subfigure}[t]{.49\textwidth} 
        \centering
        \includegraphics[width=\linewidth]{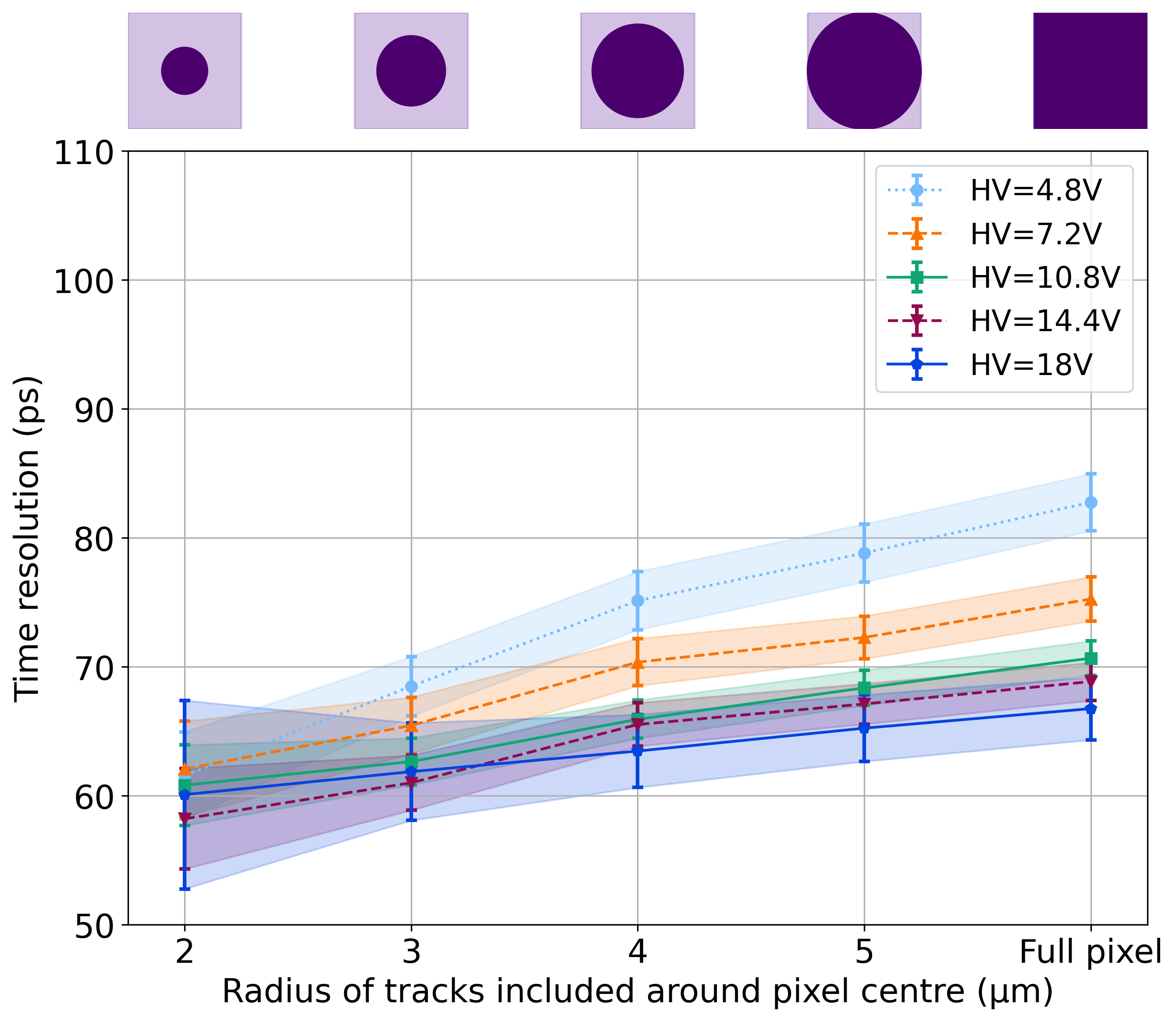}
        \caption{}
        \label{fig:3sigma_rms_cuts_corr_AC_merged_HV}
    \end{subfigure}
    \caption{The time resolution for different in-pixel selections of AC-coupled sensors at all bias voltages before (a) and after (b) a time walk-like correction. Time resolution is determined by quadratically subtracting the time resolution of the LGAD used as a time reference from the 3$\sigma$ standard deviation of the time residuals.}
    \label{fig:3sigma_rms_vsub_AC_merged_HV}
\end{figure}

Figure~\ref{fig:3sigma_rms_cuts_corr_AC_merged_HV} confirms that, as in case of the DC-coupled sensor, a smaller radial selections allows for a better timing resolution, as charge collection is dominated by drift under the collection diode. 

The results indicate that the input capacitance seen by the front-end of the AC-coupled sensor is higher, resulting in a reduced signal-to-noise ratio, as also indicated by the noise measurements. This, in turn, leads to a larger jitter contribution to the time resolution compared to the DC-coupled configuration.

\section{Conclusions}
This work presents the performance of analog pixel test structures (APTS) fabricated using the \SI{65}{\nm} TPSCo CMOS imaging technology, with a particular focus on radiation hardness and a comparison between different sensor coupling schemes to the front-end electronics.

The APTS-OA sensor was integrated into a beam telescope to evaluate its time resolution, charge collection, and detection efficiency using minimum ionizing particles. The DC-coupled sensors exhibited excellent radiation tolerance, maintaining consistent timing performance up to a fluence of \SI{d14}{\NIEL}, and achieving detection efficiencies exceeding 99\% at a substrate voltage of \SI{-4.8}{\volt} with a threshold of \SI{100}{\ele}.

AC-coupled sensors, similarly, showed a good operational margin, with efficiencies well above 99\% for clusterization thresholds lower than \SI{150}{e^-}. 
AC coupling allows the sensor to be reverse biased beyond the limits of the DC-coupled version, leading to observable improvements in time resolution up to 18 V. However, due to the lower signal amplitude introduced by AC coupling, the signal-to-noise ratio is smaller than in the DC configuration, making the jitter contribution more significant. At high reverse bias, the AC-coupled version reaches a time resolution comparable to the DC-coupled one. This demonstrates the viability of both approaches, while also highlighting the potential benefit of combining the small capacitance of the DC-coupled design with the high reverse bias capability of the AC-coupled configuration to achieve further improvements in time resolution.

These results highlight the potential of the \SI{65}{\nm} CMOS imaging process, particularly with regard to radiation hardness and high-precision timing capabilities. Monolithic active pixel sensors based on this technology are strong candidates for future high-energy physics vertex and timing detectors
but also for broader applications requiring precise timing.

\printindex
\newenvironment{acknowledgement}{\relax}{\relax}
\begin{acknowledgement}
\section*{Acknowledgements}
This project has received funding from the European Union’s Horizon Europe research and innovation programme under grant agreement No 101057511.
Funding sources: (1) Project “Dipartimenti di eccellenza” at the Dept of Physics, University of Turin, funded by Italian MIUR.
(2) The National Research Foundation of Korea (NRF) grant funded by the Korean government (MSIT) (No. 2022R1A6A3A03054907)
Project  2022LJT55R, Concession Decree No. 104 of 02.02.2022 adopted by the Italian Ministry of University and Research, cup D53D23002810006.
The measurements leading to these results have been performed at the SPS Test Beam Facility at CERN (Switzerland). 
We would like to thank the coordinators at CERN for their valuable support of these test beam measurements and for the excellent test beam environment.
ITS3 R\&D and construction is supported by several ITS3 project grants including LM2023040 of the Ministry of Education, Youth, and Sports of the Czech Republic and Suranaree University of Technology, the National Science and Technology Development Agency (JRA-CO-2563-12905-TH, P-2050706), and the NSRF via the Program Management Unit for Human Resources \& Institutional Development, Research and Innovation (PMU-B B47G670091) of Thailand.    
\end{acknowledgement}
%

\bibliographystyle{utphys}
\bibliography{references}

\newpage
\appendix



\section{Statistics acquired.}
\label{apx:statistics}

\begin{table}[ht]
\centering
\begin{tabular}{r|cccc}
DUT & \multicolumn{4}{c}{\begin{tabular}[c]{@{}c@{}}Irradiated\\ \SI{d14}{\NIEL}\end{tabular}} \\ \hline
$V_\text{sub}$ & \SI{-1.2}{\volt} & \SI{-2.4}{\volt} & \SI{-3.6}{\volt} & \SI{-4.8}{\volt} \\ \hline
All events & 70,000 & 70,000 & 90,000 & 735,000 \\
Coincident events & 3,271 & 3,214 & 3,412 & 37,618
\end{tabular}
\caption{Number of events}
\label{tab:statistics_1e14}
\end{table}

\begin{table}[!ht]
\centering
\begin{tabular}{r|cccc}
DUT & \multicolumn{4}{c}{\begin{tabular}[c]{@{}c@{}}Irradiated\\ \SI{d15}{\NIEL}\end{tabular}} \\ \hline
$V_\text{sub}$ & \SI{-1.2}{\volt} & \SI{-2.4}{\volt} & \SI{-3.6}{\volt} & \SI{-4.8}{\volt} \\ \hline
All events & 670,000 & 60,000 & 60,000 & 830,000 \\
Coincident events & 32,107 & 3,129 & 3,180 & 34,111
\end{tabular}
\caption{}
\label{tab:statistics_1e15}
\end{table}

\begin{table}[!ht]
\centering
\begin{tabular}{r|ccccc}
DUT & \multicolumn{4}{c}{AC-coupled} \\ \hline
$V_\text{HV}$ & \SI{4.8}{\volt} & \SI{7.2}{\volt} & \SI{10.8}{\volt} & \SI{14.4}{\volt} & \SI{18.0}{\volt} \\ \hline
All events & 210,000 & 220,000 & 346,000 & 210,000 & 70,000 \\
Coincident events & 10,356 & 11,077 & 17,215 & 10,800 & 3,549
\end{tabular}
\caption{}
\label{tab:statistics_ac}
\end{table}

The statistics of the total events triggered by the beam is reduced to about 5\% due to the large acceptance of the trigger plane, the signal extraction method cut, and the clusterization threshold.

\newpage
\section{X-ray spectra.}
\label{apx:xray-spectra}
\begin{figure}[ht]
    \centering
    \begin{subfigure}[t]{.49\textwidth}
        \centering
        \includegraphics[width=\linewidth]{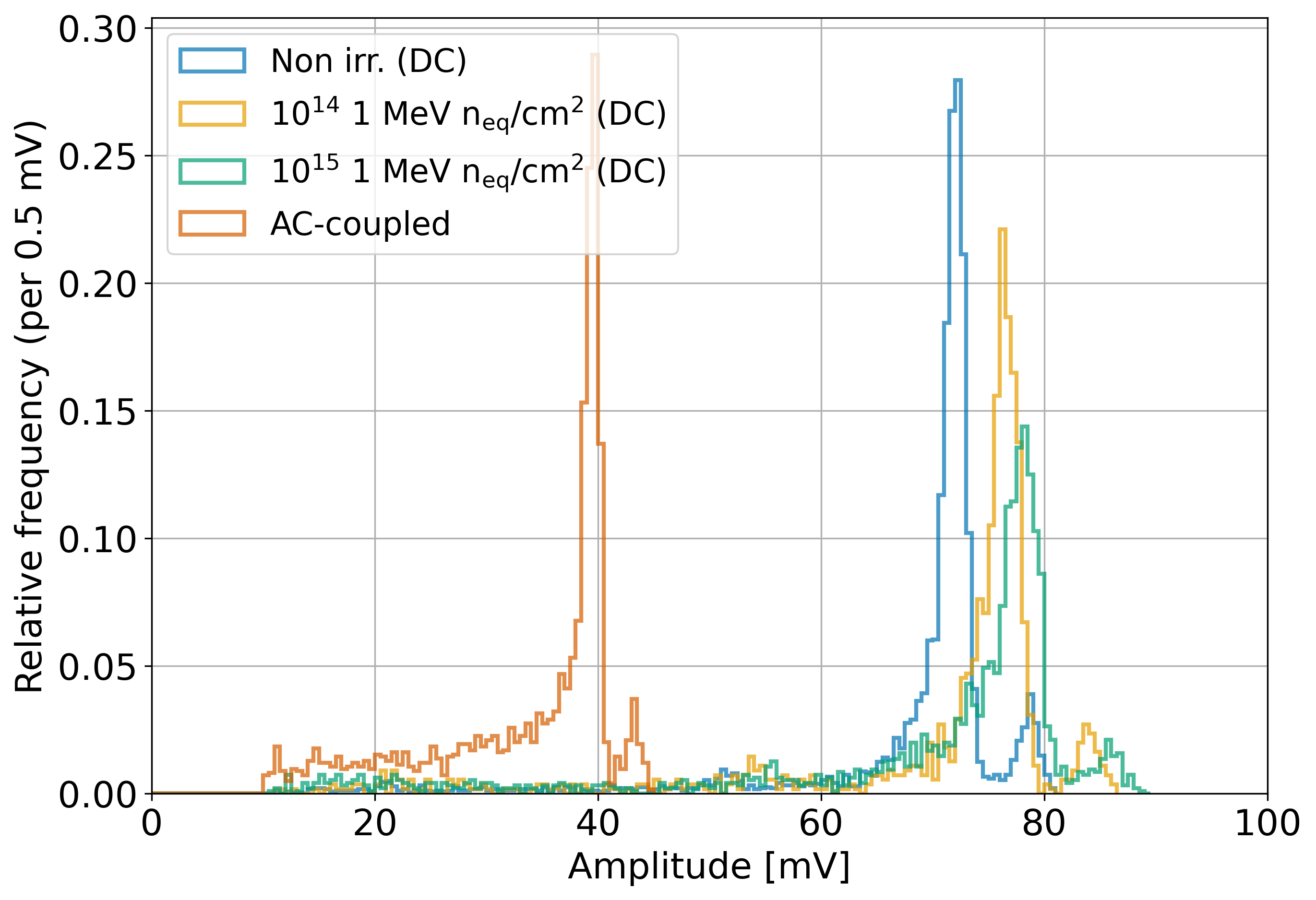}
        \caption{}
        \label{fig:x-ray_mv}
    \end{subfigure}
    \begin{subfigure}[t]{.49\textwidth}
        \centering
        \includegraphics[width=\linewidth]{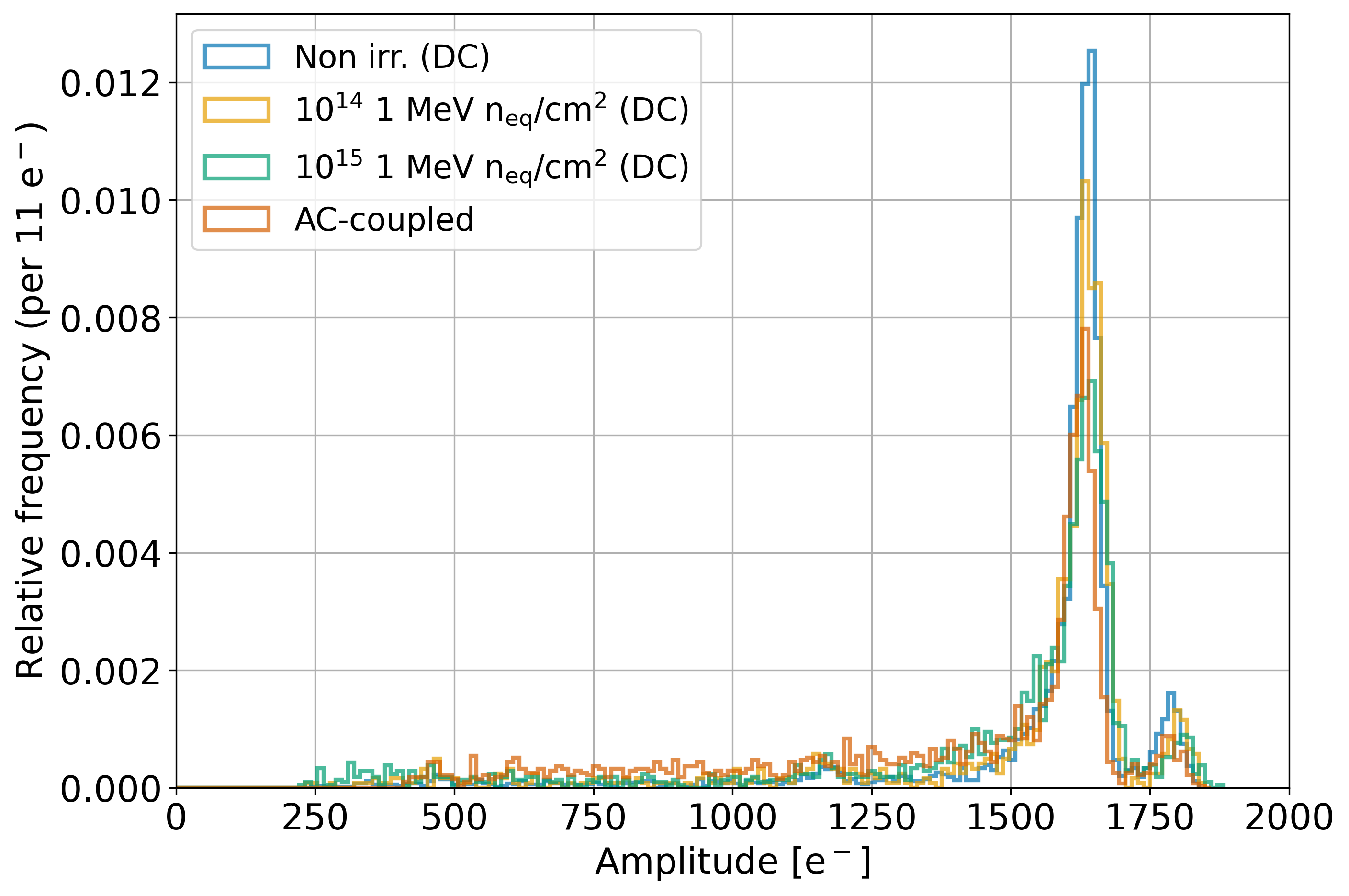}
        \caption{}
        \label{fig:x-ray_e}
    \end{subfigure}
    \caption{X-ray spectra from DC-coupled irradiated (at $V_\text{sub}=\SI{-4.8}{\volt}$) and AC-coupled sensors (at $V_\text{HV}=\SI{4.8}{\volt}$) in mV (a) and converted to electrons (b). 
    }
    \label{fig:x-ray}
\end{figure}

Figure \ref{fig:x-ray_mv} shows the X-ray spectra of irradiated and AC-coupled sensors. For the AC-coupled sensor, the $K_{\alpha}$ peak shows up at lower amplitude than the DC-coupled sensors due to a higher total capacitance. Irradiation damage decreases the energy resolution of the sensor and so broadens the $K_\alpha$ peak, as also seen in~\cite{APTSSF_2023}.
A corresponding apparent reduction in capacitance, reflected in the variation of signal amplitudes between irradiated and non-irradiated devices, in qualitative agreement with the acceptor-removal mechanism.
The conversion to electrons (Fig.\ref{fig:x-ray_e}) effectively equalizes the peak positions, differences due to the different sensor energy resolution remain.


\newpage
\section{Charge collection Landau distribution.}
\label{apx:cluster_charge}
\begin{figure}[!ht]
    \centering
    \begin{subfigure}[t]{.49\textwidth}
        \centering
        \includegraphics[width=\linewidth]{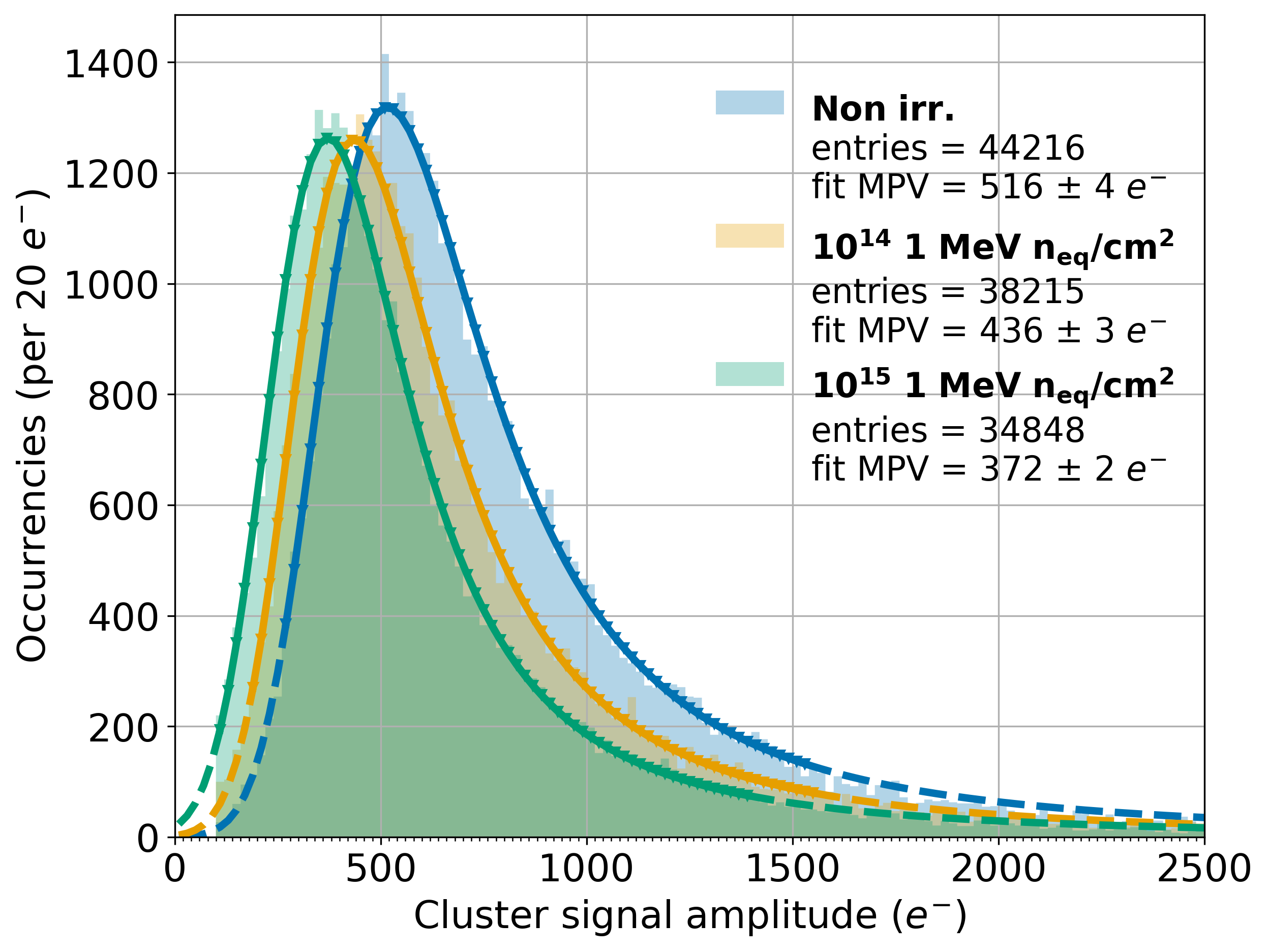}
        \caption{}
        \label{fig:landau_dist_irr}
    \end{subfigure}
    \begin{subfigure}[t]{.49\textwidth} 
        \centering
        \includegraphics[width=\linewidth]{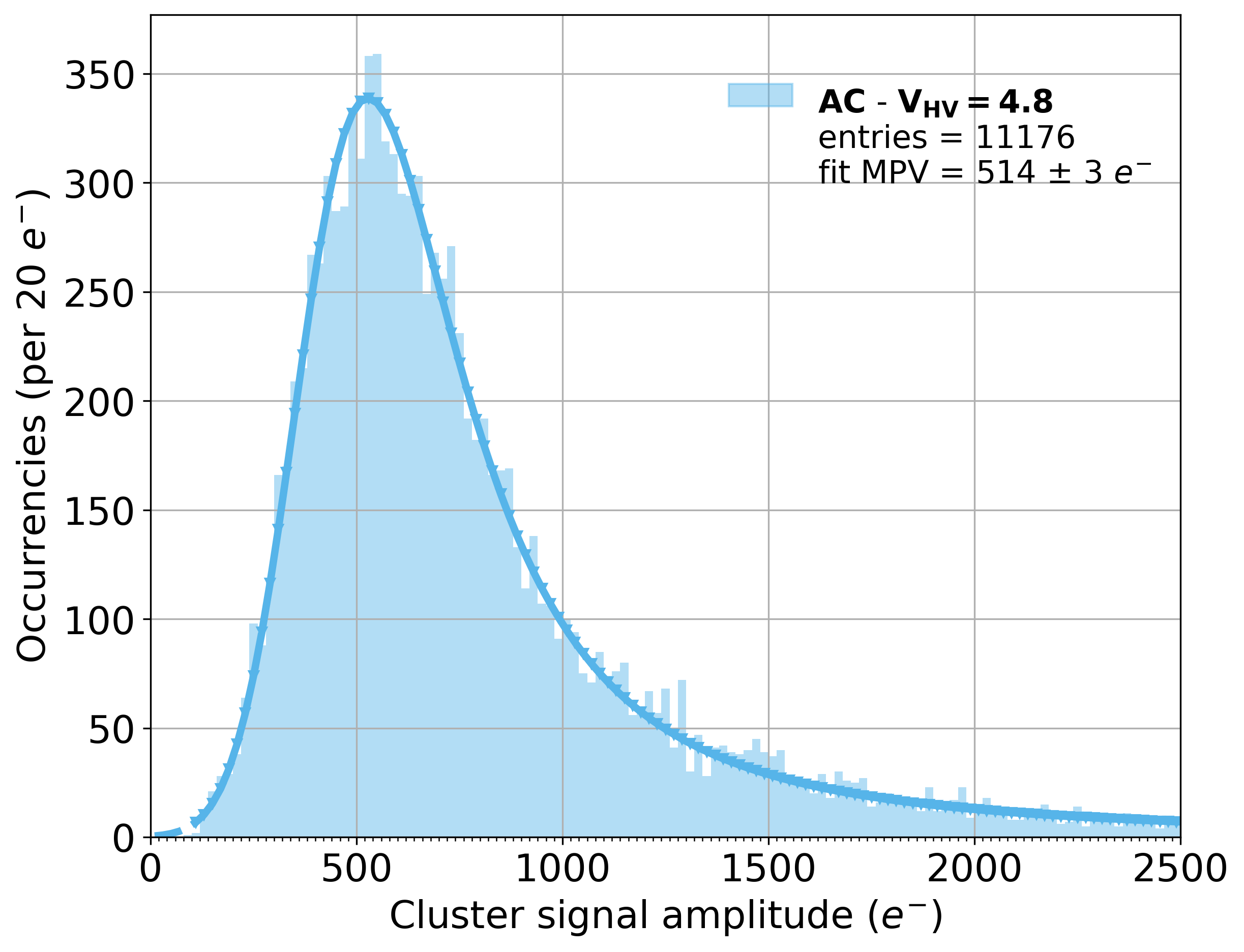}
        \caption{}
        \label{fig:landau_dist_AC_4.8}
    \end{subfigure}
    \begin{subfigure}[t]{.49\textwidth} 
        \centering
        \includegraphics[width=\linewidth]{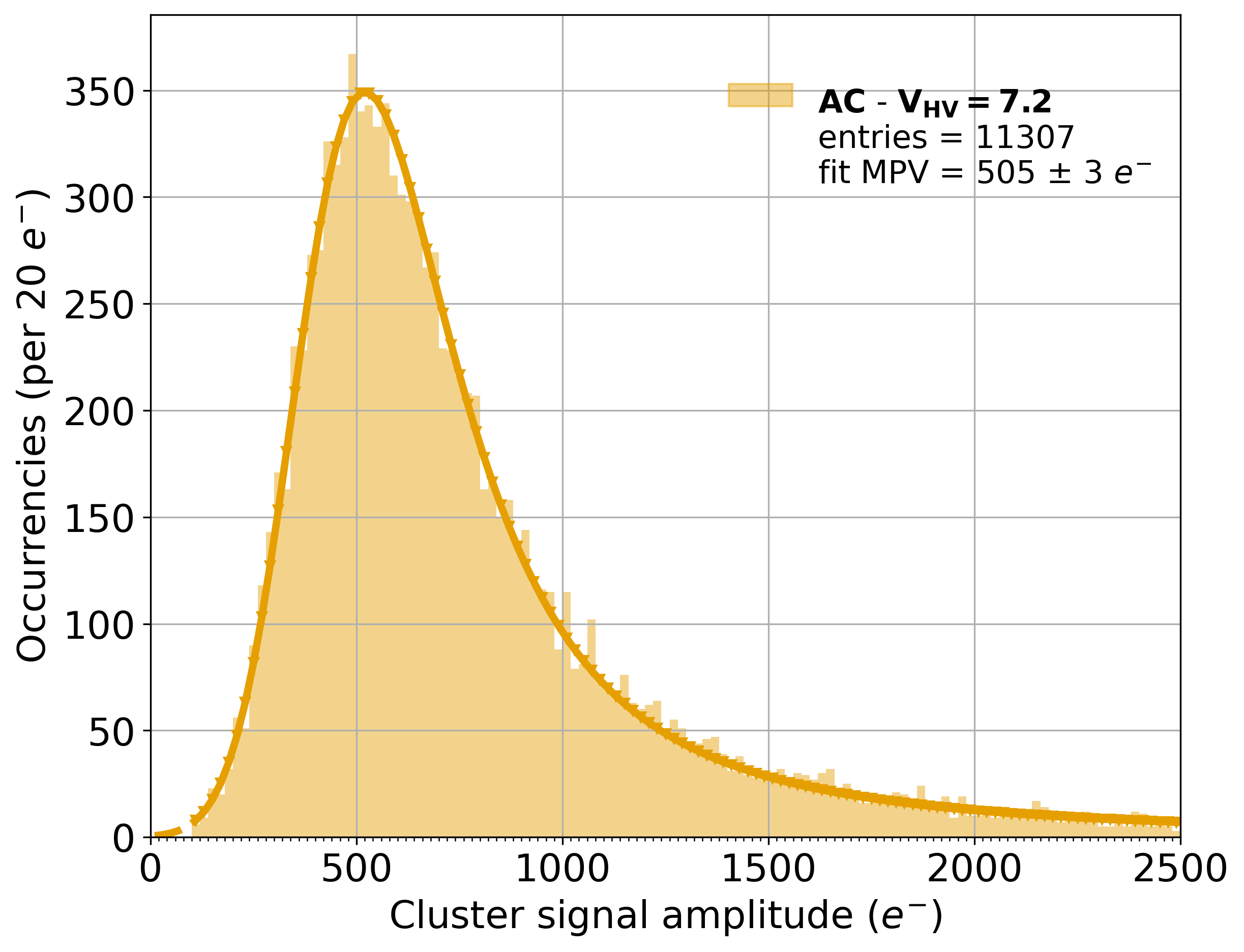}
        \caption{}
        \label{fig:landau_dist_AC_7.2}
    \end{subfigure}
    \begin{subfigure}[t]{.49\textwidth} 
        \centering
        \includegraphics[width=\linewidth]{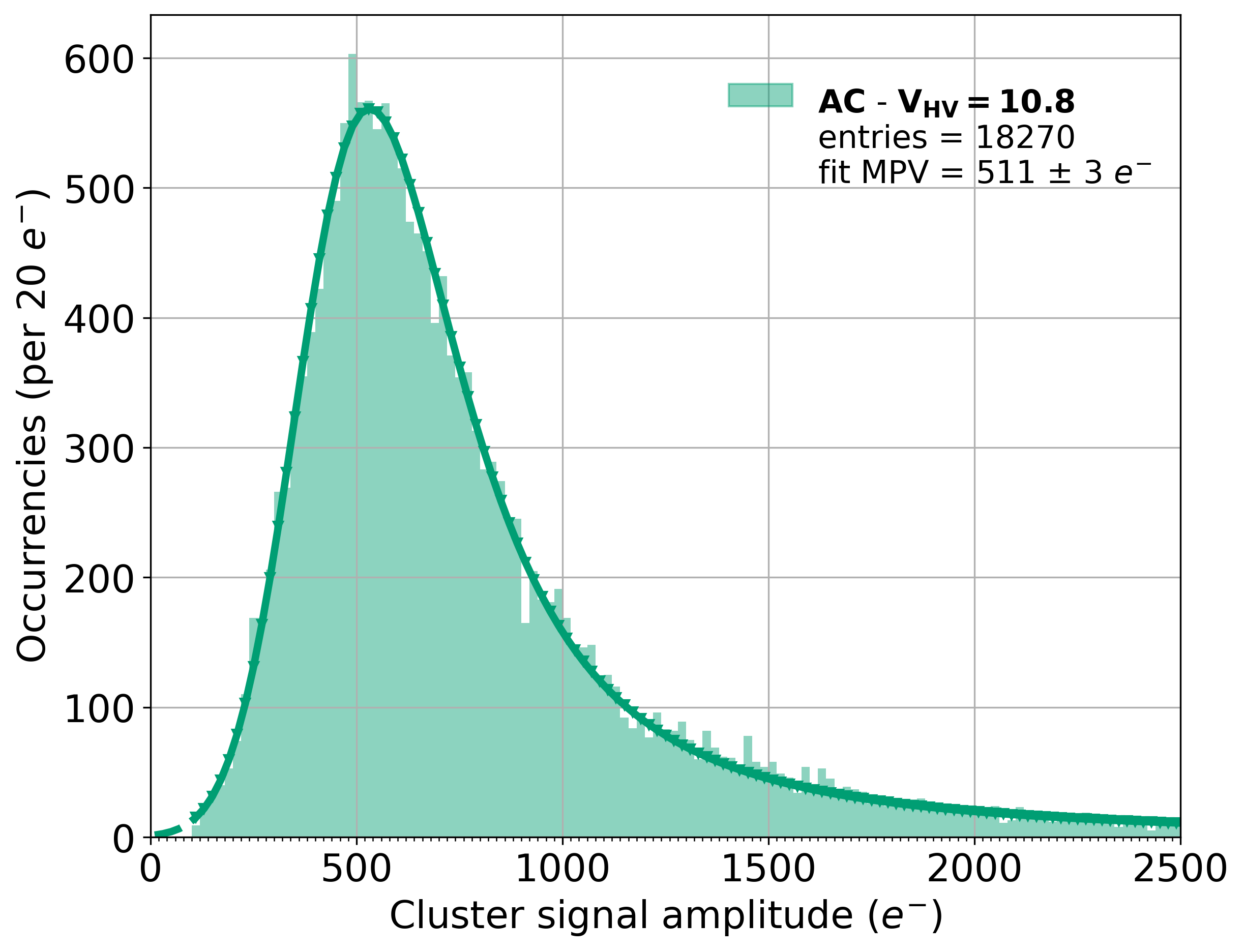}
        \caption{}
        \label{fig:landau_dist_AC_10.8}
    \end{subfigure}
    \begin{subfigure}[t]{.49\textwidth} 
        \centering
        \includegraphics[width=\linewidth]{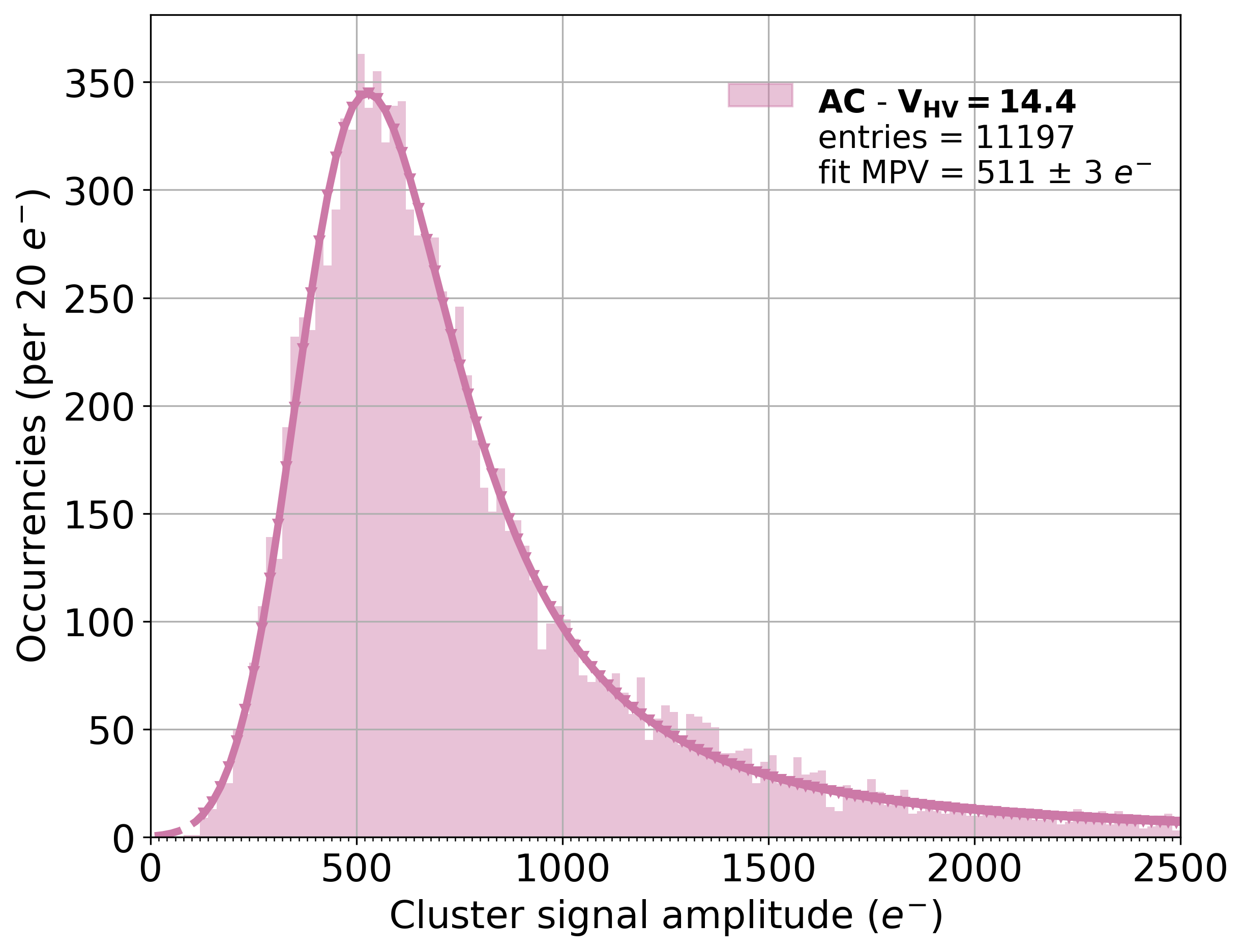}
        \caption{}
        \label{fig:landau_dist_AC_14.4}
    \end{subfigure}
    \begin{subfigure}[t]{.49\textwidth} 
        \centering
        \includegraphics[width=\linewidth]{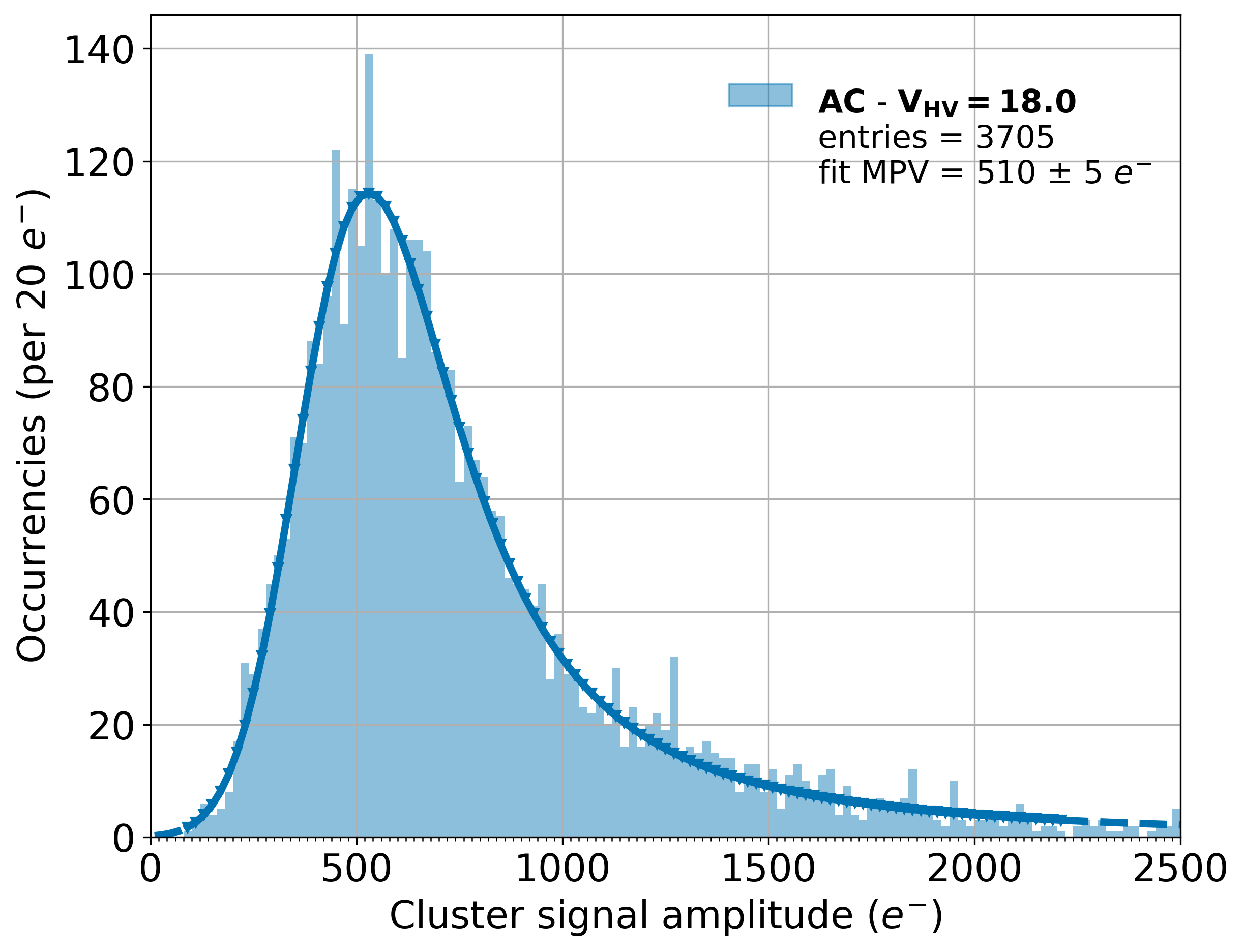}
        \caption{}
        \label{fig:landau_dist_AC_18.0}
    \end{subfigure}
    \caption{Charge collection distributions of DC-coupled irradiated sensors at $V_\text{sub}=\SI{-4.8}{\volt}$ (a) and the AC-coupled sensor at $V_{\rm{HV}}=\SI{4.8}{\volt}$, $\SI{7.2}{\volt}$, $\SI{10.8}{\volt}$, $\SI{14.4}{\volt}$ and $\SI{18}{\volt}$ (b, c, d, e and f, respectively). The data are shown in the shaded histogram with a solid line showing the result from a Landau+Gaussian fit. The dashed line shows the result of the fit outside of the fitting region.}
    \label{fig:landau_dist}
\end{figure}


\end{document}